\documentclass[paper=a4, fontsize=12pt]{article} 
\usepackage{amssymb,amsfonts,bbm,amsthm,amsmath,verbatim,graphicx}
\usepackage{xcolor,colortbl}
\usepackage[utf8]{inputenc}
\usepackage{tikz}
\usepackage[margin=1in]{geometry}
\usepackage{setspace,parskip}
\usepackage{soul}
\usepackage{natbib}
\usepackage{{booktabs}}
\usepackage{threeparttable}
\usepackage{multirow}
\usepackage{enumitem} 
\usepackage{hyperref}

\usepackage{float}
\usepackage{subcaption}

\renewcommand{\phi}{\varphi}

\usepackage[makeroom]{cancel}
\usepackage{shapepar}
\usepackage{adjustbox}
\usepackage{marginnote}
\usepackage{mathrsfs}
\usepackage{multirow}

\newcommand{\E}{\mathbbm{E}}

\usepackage[amssymb]{SIunits}
\allowdisplaybreaks

\theoremstyle{remark}

\theoremstyle{definition}

\usepackage[T1]{fontenc} 
\usepackage[english]{babel} 
\usepackage{amsmath,amsfonts,amsthm} 

\usepackage{lipsum} 

\usepackage{sectsty} 
\allsectionsfont{\normalfont\scshape}

\usepackage{fancyhdr} 
\reversemarginpar

\numberwithin{equation}{section} 

\usepackage{placeins}
\title{	
Reevaluating Causal Estimation Methods \\ with Data from a Product Release\thanks{For helpful guidance, we are grateful to Guido Imbens, Jann Spiess, Vasilis Syrgkanis, Lihua Lei, Julian Nyarko, Stefan Wager, Ramesh Johari, Jason Weitze, Lea Bottmer, David Ritzwoller, Fabio Vera, and conference and seminar participants at NABE TEC 2024, Microsoft Research, and Stanford. We are especially grateful to Muthoni Ngatia, Mary Hu, and Martin Tingley for their invaluable help in sourcing the data and enabling its release.}}
\author{
    Justin Young\thanks{Microsoft, email: youngjustin@microsoft.com} \\ 
    \and
    Eleanor W. Dillon\thanks{Microsoft Research, email: eleanor.dillon@microsoft.com}
}

\date{\today}

\begin{document}

\maketitle

\doublespacing

\setlength{\parindent}{20pt}

\begin{abstract}
    Machine learning methods estimate flexible relationships between confounders, treatments and outcomes, making unconfoundedness assumptions in causal analysis more palatable. How successful are these approaches in recovering ground truth baselines? We analyze new data covering an experimental rollout of a software product feature for one group of users and a voluntary opt-in to the same feature for another similar group. Recovering causal effects from nonrandomized data is feasible—but only with careful modeling choices. We offer best practices for more credible treatment effect estimation in modern, high-dimensional datasets and evaluate which tests provide meaningful signals about the reliability of estimated effects.
\end{abstract}

\newpage

\section{Introduction}

Many econometric techniques aim to estimate causal effects from existing, nonrandomized data. These methods are important tools to answer applied questions in economic, policy, health, and business situations where randomized experiments are infeasible or unethical. Unlike prediction problems, where estimates can eventually be compared with realized outcomes, causal effect estimates typically have no ground truth to compare against, leaving open questions about their practical reliability. Evaluation of these methods against a small set of existing benchmarks, starting with LaLonde's seminal 1986 examination of a job training program, have produced mixed results.

We introduce a new, publicly available benchmark dataset capturing a common business setting.\footnote{A repository containing the benchmark datasets, documentation, and replication materials is available at \url{https://github.com/microsoft/Reevaluating-Causal-Estimation-Methods}.} Software companies regularly push updates to existing products. These updates could introduce new features such as auto-complete in writing and coding environments, update existing features such as reorganizing toolbar layouts, or work behind the scenes to improve performance or address vulnerabilities. In many cases, these updates are added to large, complex, vintage code bases. Moreover, the final effects of these updates depend on how users interact with the product. Therefore, as with tax code changes or novel medicines, product updates designed to target one outcome will often generate additional side effects through interactions with users and existing structures that are difficult to predict in advance. 

The surest way to measure the effectiveness of updates against the target outcome, as well as identify side effects, is to conduct A/B tests that introduce the change to a random subset of existing users. However, well-designed experiments can be slow and expensive to implement and it is not always feasible or desirable to force users to accept updates. More often, a small share of existing users will opt into testing new features and software companies must do their best to disentangle the selection effects of being an early adopter from the treatment effects of the update. Our study exploits a case where Microsoft released a new feature for its Windows product in two ways at once. Windows pushed the new feature to a randomized subset of users in an A/B test while simultaneously making the new feature available to other users who could choose whether to adopt it.

 This dual roll-out creates an ideal testing ground: we measure a benchmark estimate of the effects of the new feature in the experimental sample and evaluate econometric techniques in the observational sample where uptake is nonrandom. We measure the same set of device performance outcomes in both samples, including the targeted outcome and some standard performance metrics that might exhibit side effects, and a rich set of user and device characteristics. Our large and balanced data samples improve on many of the earlier benchmarks that had limited covariate overlap between treated and untreated units in the observational sample and/or limited similarity between the observational and experimental samples. We use the richness of these data to explore the performance of a range of causal estimation techniques, including modern methods that leverage flexible machine learning, estimating both the average treatment effects (ATE) and conditional average treatment effects (CATE). Our goal is to provide practical advice for researchers: analyzing which causal methods perform well on nonrandomized data, how these methods can be used most effectively, and what checks and tests provide meaningful signals about the quality of estimated effects.

We focus on the causal effect of the new product feature on two outcomes: the targeted performance metric and one general performance metric that was also affected. In both cases, the simple difference in mean outcomes across users in the observational sample who did and did not adopt the new features tells the same qualitative story as the experimental benchmark: the new feature causes a small increase in the general performance outcome and a larger decrease in the targeted performance outcome.\footnote{The details of the performance outcomes are masked to enable the data sample to be publicly released. Positive changes do not necessarily align with ``good" outcomes.} However, in both cases these naive mean differences miss the experimentally-estimated causal effects by a meaningful margin, indicating substantial endogenous selection. For the general outcome, we are able to remove all the selection bias and recover the experimental benchmark using the available device and user characteristics and carefully deployed causal estimation methods. For the targeted outcome, none of our attempts come close to the experimental benchmark. 

Our analysis yields several bits of advice for practitioners. First, observational causal estimation methods can recover true causal effects when used carefully, but the details of estimation matter. When we follow established best practices, we find that doubly robust estimators that use flexible machine learning to capture the effects of confounders on treatment and outcome outperform simpler methods like linear regression. However, this flexibility can become a liability when those best practices are not followed. Most machine learning models have hyperparameters that shape the estimation to prevent overfitting in finite samples.\footnote{Examples include the free parameters that determine the degree of regularization in LASSO regression or the maximum depth of trees in a random forest model.} These hyperparameters should be tuned to each data sample using sample splitting to measure out-of-sample performance. When we skip this preliminary step, doubly robust methods can deliver estimated causal effects from the observational sample that are as biased as the unconditional means. In other words, design choices during estimation are equally important as modeling choices like identifying confounders. 

To our knowledge, this is the first real-world application that quantifies the importance of precisely estimating the effects of confounders on treatments and outcomes, supplementing the simulated illustrations in \citet{bach2024, cherono2018dml}. We also reinforce the importance of trimming the observational sample to remove ``always treated" or ``never treated" types \citep{DehejiaWahba1999,crump2009dealing,xuimbens2025} and averaging across estimates to reduce model uncertainty \citep{breiman1996bagging,ritzwoller2024reproducible}. Each of these best practices describe concrete, verifiable steps that can be executed with confidence in applied settings, even those without ground truth benchmarks.

The methods we use to estimate causal treatment effects from nonrandomized data all rely on an unconfoundedness assumption \citep{rosenbaum1983central}: that we have measured enough covariates to control for any dependence between treatment assignment and the realization of outcomes. Our second guidance for practitioners is to use multiple metrics to assess the plausibility of this core assumption, which cannot be tested directly. The new feature we study has, reassuringly, a very strong effect on the targeted performance metric. In the observational sample, we overstate the true treatment effect even using our best methods, estimating declines in the probability of the outcome of 54\% rather than the true 43\%, indicating that we may be missing some key confounders. We conduct sensitivity analysis following \citet{chernozhukov2022}, commonly used as one check of unconfoundedness. Our estimated treatment effects on the targeted outcome strongly ``pass" the most common test: they are unlikely to be pushed all the way to zero by a plausible unobserved confounder\footnote{We benchmark the strength of the unobserved confounder against our most powerful observed confounder.} because the effect is too strong. When only the existence and direction of the effect matters, this check may provide sufficient reassurance. However, when magnitudes matter, for example when treatment is costly, we recommend additional checks including considering the fit of the preliminary prediction models (our available confounders explain almost none of the variation in the targeted outcome) and additional sensitivity checks against other counterfactuals besides zero, for example the cost of treatment.

Finally, we use the scale and richness of our data to conduct a novel real-world test of conditional average treatment effect estimators. We validate the DR-score \citep{chernozhukov2024applied} method to detect heterogeneous treatment effects. In both the experimental and observational samples, this test suggests meaningful variation in treatment effects on the targeted outcome but not on the general outcome. Our preferred CATE estimators also identify the same user and device features driving the heterogeneity for the targeted outcome in both the experimental and observational samples, for example devices that are used more frequently experience larger declines in the targeted outcome, although every sub-group displays the same persistent selection bias when comparing estimates in the nonrandomized sample to the experimental benchmark.

Our study joins a series of validation papers using other paired observational-experimental datasets, starting with \citet{LaLonde1986} and \citet{FriedlanderRobins1995}. \citet{Heckman1997,Heckman1998} identified key conditions for credible observational estimates—such as detailed longitudinal covariate data and geographically proximate comparison groups—to mitigate selection biases. The fundamental importance of comparability across samples was further illustrated in a series of papers exploring the potential and limitations of matching methods \citep{DehejiaWahba1999,DehejiaWahba2002,HotzImbensMortimer2005,SmithTodd2005}. \citet{xuimbens2025} reviews these earlier studies and re-analyses LaLonde's sample, connecting the earlier matching studies to the general principle of overlap \citep{rubin1974estimating}. Our new data sample, which is larger than earlier benchmarks with greater overlap, allows us to reinforce some of these earlier findings with more precision, as well as adding novel insights on the importance of design-stage choices and exploring sub-group conditional average treatment effect estimation.

In section 2, we introduce our paired datasets. In section 3, we outline our formal set-up and the various estimators we consider. In section 4, we discuss our main results using the general outcome in our new data sample as the benchmark and illustrate the importance of estimation design choices. Section 5 re-tests our core methodologies using the targeted outcome in our new data sample as well as LaLonde's job training sample. Section 6 evaluates tests of unconfoundedness and heterogeneous treatment effects, while section 7 concludes.

\section{Introducing a New Dataset}

We evaluate the capacities of causal analysis methods using a new data set describing the performance and characteristics of a sample of computers running Windows. The treatment of interest is a targeted indicator of whether each device had a new feature enabled. The details of the studied feature are masked so the data can be released publicly. Examples of common feature updates in Windows include changes to the layout of the desktop and taskbar, new included applications or capacities, and changes in security procedures. The data cover the first two weeks of December, 2022. At that point, the new feature had been available to Windows users for some months, but users had to manually switch on the feature through settings and very few had done so. 

To evaluate the feature before broader release, Microsoft ran an experiment on a sample of existing, privately owned devices running Windows that had not previously turned on the feature. Microsoft switched on this feature for these two weeks for a random subset of this experimental sample. The remaining experimental devices did not have the option to turn on the feature for the two weeks of the experiment. We pair this experimental data sample with an observational sample covering the same two weeks in December. The devices in the observed sample were selected on all the same characteristics as the experimental sample except whether they had turned on the new feature. Some devices in this observed sample had already manually opted in to enabling the new feature, and kept the new feature on over the two weeks, while others had not. We therefore have two broadly similar samples: an experimental sample with exogenously assigned treatment assignment, and an observational sample where treatment is endogenously selected by device users.

Table \ref{tab:data_features} describes the full set of features in the data sample. In addition to the new feature indicator, we measure the binary targeted device performance outcome, one continuous general performance metric that is not specifically targeted by the new feature, and a variety of features describing the technical specifications and patterns of use of the devices included in the sample. The two device performance metrics are not closely related; a $t$-test cannot reject equal means of the general performance metric across the two classes of the targeted outcome. All features are observed at the device level (e.g. a laptop); characteristics of the typical user for each device are inferred from use patterns. While we do not observe lagged outcomes in this sample, several of the included features summarize patterns of past user behavior that could drive endogenous selection into turning on new features. To prepare these data for public release we have masked some features (for example scrambling geographic segment and anonymizing device manufacturer), combined some rare entries in categorical features into an ``other'' category, and truncated continuous variables at the 97th percentile.

The observational and experimental samples are well balanced across all features, as shown in Appendix Table \ref{tab:between_sample_smd}. Treatment is rare in both samples, with only 4.5\% of devices opting into the new feature in the observational sample and only 2.6\% of the experimental sample selected for treatment. Appendix Table \ref{tab:within_sample_smd} confirms that the experimental sample passes the standard balance test for randomization, with only one pre-treatment feature barely crossing the threshold of more than a 0.1 standard deviation difference in means between the treated and control devices.\footnote{Note that the observational and experimental samples are different on one dimension; the experimental sample was drawn only from devices whose users had not yet opted to the new feature, i.e., the untreated segment of the observed sample. However, opting into the new feature prior to the experiment was sufficiently rare and, as we show in the following section, imperfectly predictable so that there are no significant differences in the samples overall.}

Appendix Table \ref{tab:within_sample_smd} also shows that the observational data sample captures some important elements of endogenous selection along two dimensions. First, devices that enabled the new feature were used more frequently (more total seconds using the device and using the two most common internet browsers) and intensively (more likely to follow the patterns of highly engaged users) in the weeks before the observation period. Device usage likely affects treatment update mechanically because users who opened their devices more often had more chances to see and accept the optional update. Engagement is also a proxy for experience and expertise, which may make users more eager and confident when considering new features. Usage also likely affects performance outcomes through both channels. More frequent and intense use mechanically creates more opportunities for performance issues. More experienced users could experienced better device performance by making good choices, but could also reduce performance if they push the limits of device capabilities. Endogenous adoption of the new feature is also correlated with several device manufacturers and specifications, which could be additional proxies for the type and intensity of activities typically done on the device.

\section{Methodology and Estimation} \label{sec:methodology}

In this section we introduce precise definitions of the treatment effects of interest. We evaluate multiple estimators of these causal treatment effects, all of which rely on two key assumptions for identification: unconfoundedness and overlap.

\subsection{Estimands of Interest} 

Using the potential outcomes framework \citep{rubin1974estimating}, we can define two outcomes for each unit $i=1,\ldots,N$. $Y_i(1)$ represents the outcome of the individual had she activated the new feature, while $Y_i(0)$ represents her outcome had she not. The causal effect for unit $i$ is the difference in these potential outcomes. Due to the "fundamental problem of causal inference" \citep{holland1986}, we do not observe both potential outcomes and hence this individual causal effect cannot be directly calculated. Let $D_i \in \{0,1\}$ denote the treatment status, i.e. whether individual $i$ participates in the new feature. We observe the outcome given by $Y_i(D_i) = D_i Y_i(1) + (1 - D_i) Y_i(0)$. 

One feasible object of interest is the average treatment effect (ATE), the average effect of treatment across the population. 
\begin{equation}
ATE = E[Y_i(1) - Y_i(0)].
\end{equation}
Another object of interest might be the average treatment effect on the treated, 
\begin{equation}
 ATT = E[Y_i(1) - Y_i(0) \mid D_i=1].
\end{equation}
If the set of observations that are treated in the data used for estimation is representative of all units that are ever likely to be treated, then the average treatment effect on the treated is the most policy-relevant estimand. If decision makers would like to consider the impact of a universal treatment policy, the ATE is more appropriate. 

If all units respond in the same way to treatment (homogeneous treatment effects), then ATE=ATT. In contrast, if different units respond differently to treatment, then the group of units treated in the estimation data is only one sub-group of interest. We can also consider the more general case of a conditional average treatment effect (CATE), 
\begin{equation}
\tau(x) = E[Y_i(1) - Y_i(0) \mid X_i = x],
\end{equation}
which captures heterogeneity across any set of covariates $X$. Estimating the CATE allows researchers to examine subgroup effects and personalize policy recommendations.

\subsection{Identifying Assumptions}

To identify these estimands, we rely on two core assumptions.

\paragraph{Unconfoundedness: } Also known as selection on observables, this key identifying assumption requires that treatment assignment is independent of the potential outcomes conditional on observed covariates $X$ \citep{rosenbaum1983central}:
\begin{equation}
Y(0), Y(1) \perp D \mid X.
\end{equation}
This assumption ensures that conditional on $X$, treated and untreated units are comparable. A weaker version suffices for ATT estimation:
\begin{equation}
Y(0) \perp D \mid X.
\end{equation}
Since unconfoundedness is fundamentally untestable, researchers often conduct tests such as sensitivity analyses to examine its plausibility \citep{abadie2006large}, as we discuss further in section \ref{sec:sensitivity}.

\paragraph{Overlap:} This assumption ensures that all units have a nonzero probability of receiving treatment:
\begin{equation}
0 < P(D = 1 \mid X) < 1 \quad \forall X.
\end{equation}
Overlap ensures that it is possible to use untreated units to construct an estimate of the counterfactual outcome for each treated unit, and vice versa, without extrapolation. Extrapolation can lead to meaningful bias \citep{xuimbens2025}. For the ATE, this \textit{common support} condition must hold across the entire covariate distribution. For the ATT, it is only necessary that treated units could plausibly have received the control. Formally, this requires $P(D = 1 \mid X) < 1$ for all $X$, so that every treated unit has a corresponding control counterpart. Areas of the covariate space populated only by control units are allowed, since they are not needed to identify the ATT.

Fortunately, this condition is easy to assess using estimates of the propensity to be treated, $e(X) \equiv P(D = 1 \mid X)$. If overlap is not met for the full data sample, researchers can estimate treatment effects on a trimmed subsample of data for which overlap is satisfied. For all of the estimators we consider, we begin with this preliminary step, following
the optimal trimming rule designed by \citet{crump2009dealing} where the trimmed sample $\mathbb{A}$ is selected from the full sample $\mathbb{X}$
\[
\mathbb{A}^* = \{x \in \mathbb{X} \mid \alpha \leq e(X) \leq 1 - \alpha\}
\]
\noindent and \( \alpha \) is chosen to minimize the asymptotic variance of treatment effect estimates. 

After trimming, we can achieve an unbiased estimate of a slightly different estimand: the average treatment effect within the region with strong covariate overlap. This shift reduces sensitivity to model misspecification and improves internal validity at the potential cost of some external validity. This overlap ATE may vary meaningfully from the population ATE, particularly in settings such as ours where treatment is rare. Overall, an unbiased estimate of this subgroup effect is likely preferable to an estimate of a population ATE that is biased in an unknown direction, but researchers should keep the distinction in mind when interpreting their results.

\subsection{Methods}

To estimate treatment effects, we employ a range of estimators, each employing different strategies to control for confounding. Modern econometric estimators leverage advancements from machine learning to flexibly estimate \textit{nuisance} functions and maximize the predictive power of the set of observed confounders. Specifically, most of the methods we evaluate require estimates of the conditional mean outcome function $\mu_d(X) \equiv \E[ Y \mid D=d, X]$, the propensity score $e(X)$, or both. 

\subsubsection{Outcome Modeling}

One strategy for estimating causal effects under unconfoundedness is regression adjustment, or outcome modeling, which aims to estimate the conditional mean of potential outcomes given covariates. Formally, for a binary treatment, this involves modeling two conditional response functions, $\mu_0(X)$ and $\mu_1(X)$. The simplest and most widely used implementation is linear regression, which assumes each conditional response surface is linear, $\mu_0(x) = X'\beta$), and additive, $\mu_1(x) = \mu_0(x) + \tau$ and $\mu_0(x) = X'\beta$. When these conditions hold, the coefficient on the treatment indicator in a regression of the form $Y_i = \alpha + \tau D_i + X_i'\beta + \varepsilon_i$ consistently estimates the average treatment effect \citep{angristpischke2009}. The linear regression model is simple to implement, easily interpretable, and computationally efficient. However, it relies on strong assumptions and may be biased when these conditions do not hold.

To allow for treatment effect heterogeneity or better finite-sample performance, researchers may instead model $\mu_0(X)$ and $\mu_1(X)$ separately. This two-model approach, often referred to as the Oaxaca-Blinder estimator \citep{kline2011}, estimates treatment effects as the difference in these two predicted outcomes for the ATE, 

\begin{equation}
\hat{\tau}^{ATE} = \frac{1}{N} \sum_{i=1}^{N} \left[ \hat{\mu}_1(X_i) - \hat{\mu}_0(X_i) \right],
\end{equation}
\noindent or as the difference between realized and predicted counterfactual outcomes for the ATT,
\begin{equation}
\hat{\tau}^{ATT} = \frac{1}{N_1} \sum_{i: D_i=1} \left[ Y_i - \hat{\mu}_0(X_i) \right],
\end{equation}
where $N_1$ denotes the number of treated units.

This more flexible outcome modeling can be efficient when the outcome models are correctly specified \citep{hahn1998role}. However, because outcome modeling relies on the correct specification of the response surface(s), errors in the functional form can lead to significant bias \citep{guido2004}. This motivates using semiparametric or nonparametric methods to estimate these functions, including machine learning approaches which offer greater flexibility in high-dimensional or nonlinear settings \citep{athey2019generalized}.


\subsubsection{Propensity Score Modeling}

Another class of estimators relies on propensity score modeling. Researchers can use estimated propensity scores, $e(X_i)$, to construct counterfactuals for treated units by pairing them with control units that had a similar predicted probability of treatment \citep{rosenbaum1983central, abadie2016matching}. This propensity score matching reduces potentially high-dimensional covariate sets into a single, salient statistic, balancing treatment and control groups while maintaining computational feasibility. Under unconfoundedness and overlap, the propensity score is a sufficient statistic for deconfounding \citep{rosenbaum1983central}. 

Let $\mathcal{M}_K(i)$ denote the indices of the $K$ nearest neighbors (in propensity score) of unit $i$ from the opposite treatment group. The matching estimators are:

\begin{equation}
\hat{\tau}_{ATE} = \frac{1}{N} \sum_{i=1}^{N} (2D_i - 1) \left( Y_i - \frac{1}{K} \sum_{j \in \mathcal{M}_K(i)} Y_j \right)
\end{equation}

\begin{equation}
\hat{\tau}_{ATT} = \frac{1}{N_1} \sum_{i: D_i = 1}\left( Y_i - \frac{1}{K} \sum_{j \in \mathcal{M}_K(i)} Y_j \right).
\end{equation}

While intuitive, PSM relies on correctly specifying the propensity score model, and its performance depends on overlap between treated and untreated units. Inadequate overlap can lead to poor matches, Balance diagnostics and trimming can help mitigate these issues \citep{DehejiaWahba1999}. Matching \textit{with} replacement and caliper restrictions further improve robustness when common support is limited \citep{DehejiaWahba2002}. 

Another approach using propensity scores is inverse propensity weighting (IPW). This method creates a weighted pseudo-population of units using the estimated propensity score $\hat{e}(X_i)$ \citep{hirano2003efficient}:
\begin{equation}
\hat{\tau}^{ATE} = \frac{1}{N} \sum_{i=1}^{N} \left[ \frac{D_i Y_i}{\hat{e}(X_i)} - \frac{(1 - D_i) Y_i}{1 - \hat{e}(X_i)} \right].
\end{equation}
\begin{equation}
\hat{\tau}^{ATT} = \frac{1}{N_T} \sum_{i: D_i=1} \left[ Y_i - \sum_{j: D_j=0} \frac{\hat{e}(X_j)}{1 - \hat{e}(X_j)} Y_j \right].
\end{equation}
While IPW is unbiased under correct specification, large or unstable weights due to extreme $\hat{e}(X_i)$ values can increase variance and degrade finite-sample performance, making trimming or stabilized weights necessary in practice \citep{crump2009dealing}. Nonparametric estimation of the propensity score (e.g., via random forests or gradient boosting) is now commonly used to relax modeling assumptions \citep{li2018balancing}.

\subsubsection{Augmented Inverse Probability Weighting}
Augmented inverse probability weighting (IAPW) combines IPW and outcome modeling, producing a doubly robust estimator \citep{guido2004, glynn2009aipw}: 
\begin{equation}
\hat{\tau}^{ATE} = \frac{1}{N} \sum_{i=1}^{N} \left[ \left( \frac{D_i}{\hat{e}(X_i)} - \frac{1 - D_i}{1 - \hat{e}(X_i)} \right) (Y_i - \hat{\mu}_{D_i}(X_i)) + \hat{\mu}_1(X_i) - \hat{\mu}_0(X_i) \right].
\end{equation}
\begin{equation}
\hat{\tau}^{ATT} = \frac{1}{N_T} \left[  \sum_{i: D_i=1}\left(Y_i - \hat{\mu}_0(X_i) \right)  + \sum_{i: D_i=0} \frac{\hat{e}(X_i)}{1-\hat{e}(X_i)}\left({Y_i - \hat{\mu}_0(X_i)}\right) \right].
\end{equation}
AIPW earns its namesake as a ``doubly-robust'' estimator as it remains consistent if \textit{either} $\hat{e}(X_i)$ or $\hat{\mu}(X_i)$ is correctly specified.  This method often outperforms single model approaches in finite samples \citep{robins1995analysis, guido2004}.

\subsection{Conditional Average Treatment Effect Estimation}

For personalized policy insights and subgroup analysis, researchers are often interested in heterogeneous treatment effects $\tau(X) = \mathbb{E}[Y(1) - Y(0) \mid X]$. Our evaluation focuses on a family of estimators known as meta-learners, which reframe CATE estimation as a sequence of supervised learning problems and adapt flexible machine learning models to this causal setting \citep{kunzel2019}. Meta-learners use propensity scores and/or outcome models to construct pseudo-labels in structuring the learning task. Our primary CATE estimator is an extension of the doubly robust AIPW estimator.

The CATE DR-learner \citep{kennedy2020} constructs doubly-robust pseudo-outcomes by combining estimates of the outcome regression $\mu_D(X) = \mathbb{E}[Y \mid D, X]$ and the propensity score $e(X) = \mathbb{P}(D = 1 \mid X)$. The pseudo-label for unit $i$ is given by:
\begin{equation}
Y_i^{DR} = \left(\frac{D_i - \hat{e}(X_i)}{\hat{e}(X_i)(1 - \hat{e}(X_i))}\right)(Y_i - \hat{\mu}_{D_i}(X_i)) + \hat{\mu}_1(X_i) - \hat{\mu}_0(X_i).
\end{equation}
We fit a regression model to predict $Y_i^{DR}$ using covariates $X_i$, yielding an estimate of $\tau(X)$. Under standard regularity conditions, this CATE DR-learner is consistent and has the familiar, favorable doubly-robust property where the estimate is first-order insensitive to nuisance function estimation error. We also estimate the CATE with other meta-learners like the R- \citep{niewager2019} and X- \citep{kunzel2019} learners for comparison.

\section{Recovering Experimental Causal Effects}

We analyze the feature release and device performance data to assess how successfully each of the causal inference techniques described in the previous section can recover experimental ground truth causal effects. We begin by considering only the treatment effect of the new feature on the general, non-targeted outcome, which we consider to be the cleanest testing ground. We explore the targeted outcome, as well as additional data samples, in section \ref{sec:addl_samples}.

\subsection{Establishing Ground Truth}
\label{sec:ground_truth}

\begin{figure}[htbp]
    \centering
    \begin{subfigure}{0.45\textwidth}
    \includegraphics[width=\linewidth]{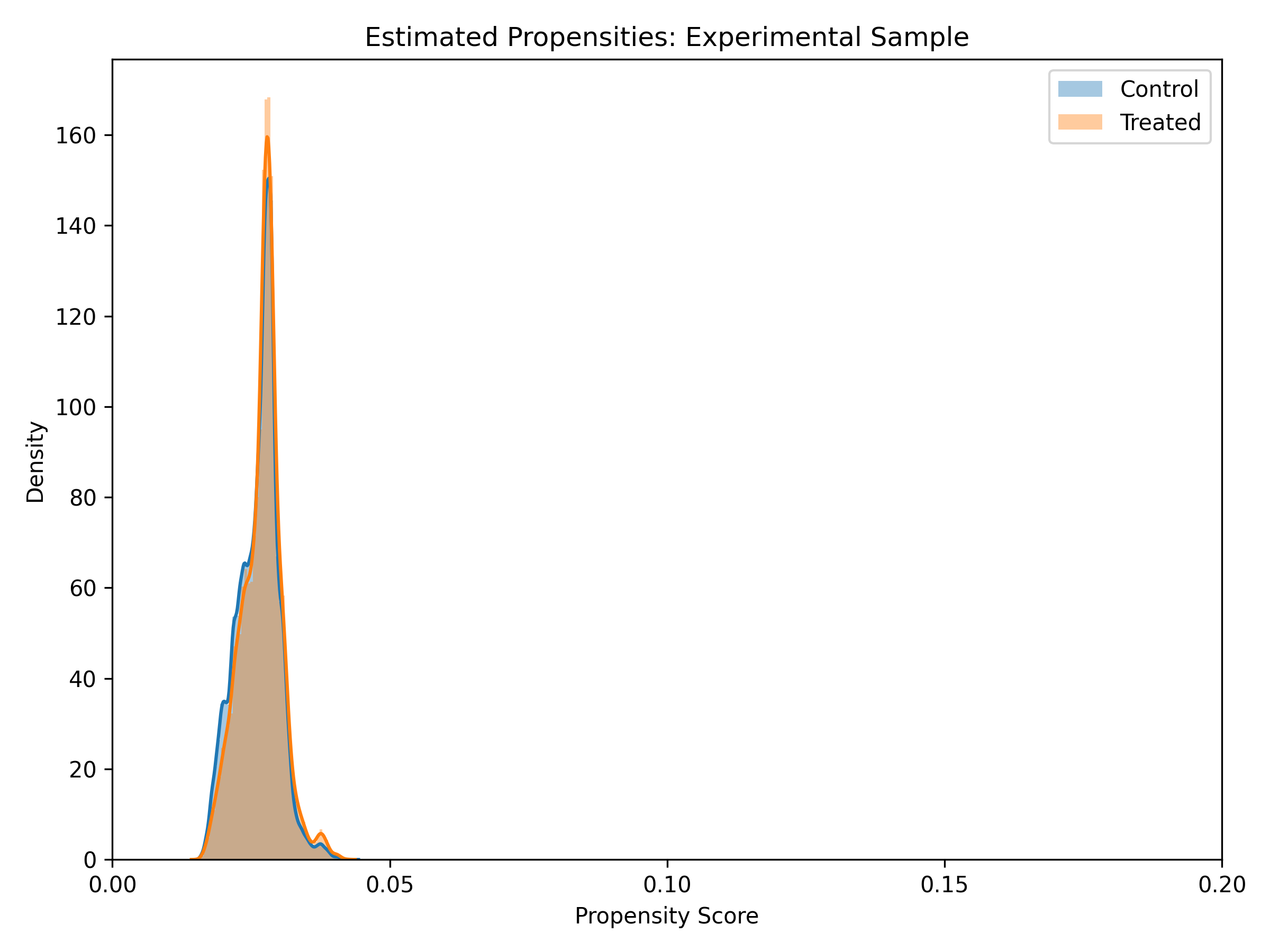}
    \caption{Experiment Sample}
    \end{subfigure}
    \begin{subfigure}{0.45\textwidth}
    \includegraphics[width=\linewidth]{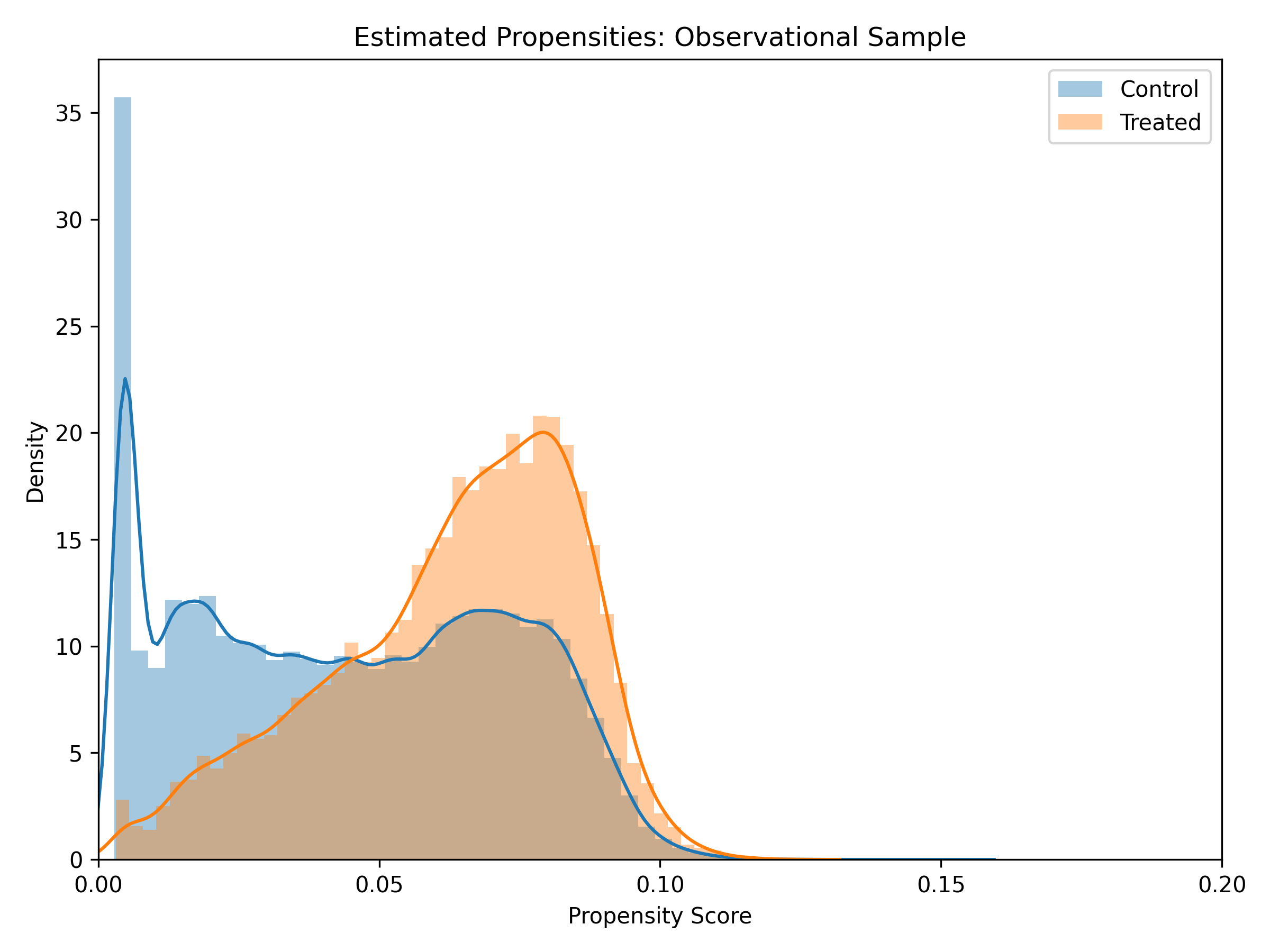}
    \caption{Observed Sample}
    \end{subfigure}
    \begin{subfigure}{0.45\textwidth}
    \includegraphics[width=\linewidth]{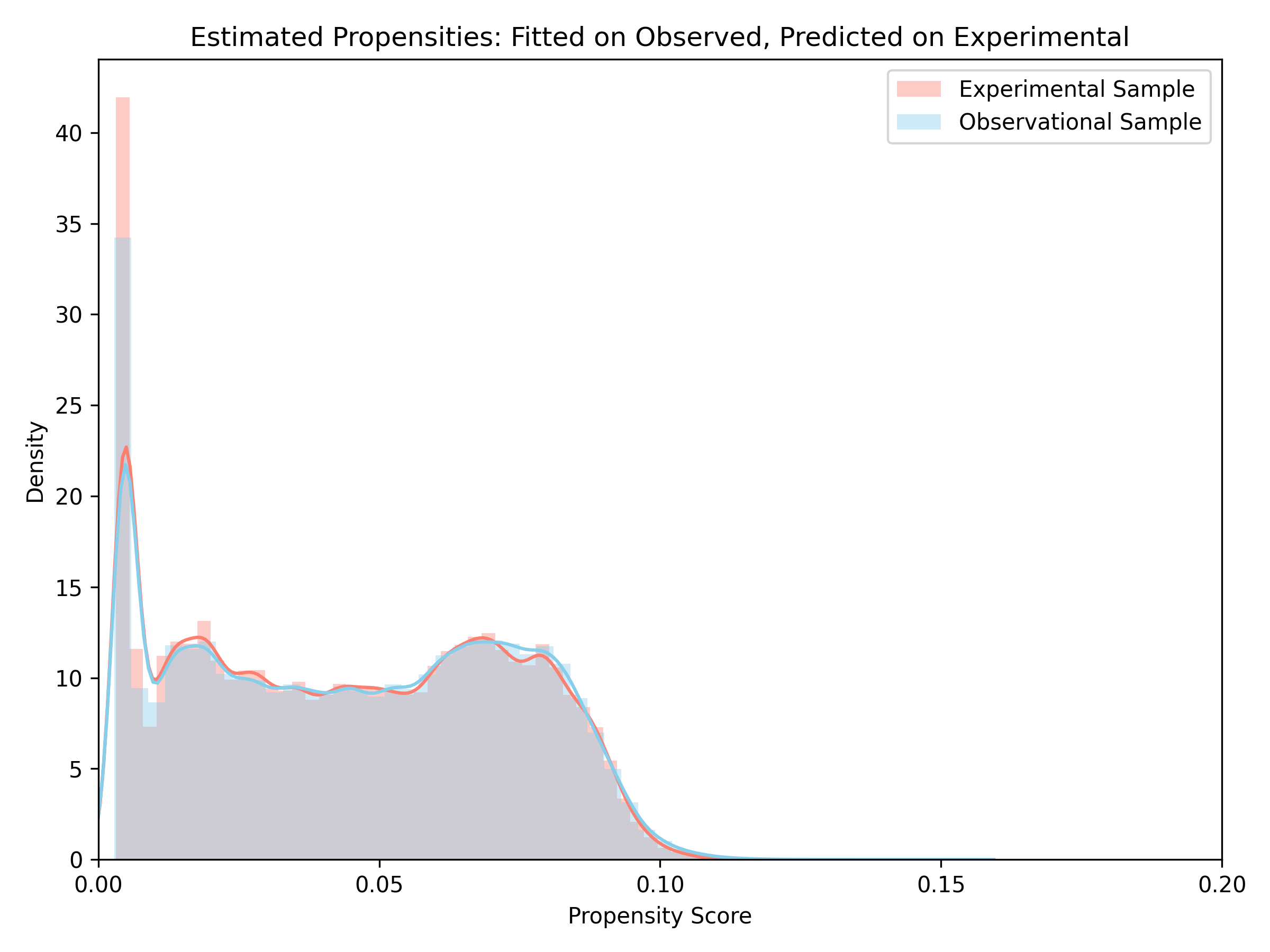}
    \caption{Observed Fit on Experiment}
    \end{subfigure}
    \caption{\textbf{Estimated Propensity Score Distributions Across Samples.} 
    Panel (a) shows the distribution of cross-fitted propensity scores estimated  on the experimental sample, reflecting the randomized assignment—most scores cluster around the overall treatment rate of 2.64\%. 
    Panel (b) shows the distribution of cross-fitted propensity score estimated on the observational sample, where greater separation between treated and untreated groups is evident, consistent with endogenous selection into treatment. 
    Panel (c) plots the observational propensity model $\hat{e}(\cdot)$ evaluated on both the entire observational and entire experimental samples. The complete distributional overlap in Panel (c), coupled with the same range of scores as seen in Panel (b), indicates that the experimental and observational samples cover comparable covariate regions, allowing trimming procedures to be applied consistently across both datasets.
    }
    \label{fig:propensity_plots}
\end{figure}

Figure~\ref{fig:propensity_plots} displays the distribution of the estimated probabilities of receiving the new product, estimated using our preferred approach of an ensemble of tuned, boosted trees with cross-fitting \citep{friedman2001greedy}. Panel (a) of Figure~\ref{fig:propensity_plots} shows limited predictive power in the experimental sample, with mass concentrated around the mean treatment rate and minimal separation between treated and untreated devices. This pattern is consistent with Appendix Table~\ref{tab:within_sample_smd} and a correctly randomized experiment. In contrast, the observational sample, in Panel (b), shows clear separation in the estimated propensities, consistent with endogenous selection into treatment.

Panel (c) of Figure~\ref{fig:propensity_plots} plots the propensity model $\hat{e}(X_i)$ estimated from the observed sample, which captures the process of endogenous selection into treatment, applied to the experimental sample. In other words, $\hat{e}(X_i)$ is learned from the observed sample and then used to predict the devices for both the observed and experimental samples. This distribution is overlaid with the full distribution of propensity scores in the observational sample (combining treated and untreated units). The range of ex ante propensities to endogenously select treatment are virtually identical in the two samples, highlighting their fundamental similarity\footnote{We test similarity feature by feature in Appendix Table~\ref{tab:between_sample_smd}.} This full overlap of the experimental and observational samples stands in contrast to some earlier benchmarks such as the \citet{LaLonde1986} data. 

To ensure an apples-to-apples comparison, we estimate the ground truth treatment effects in the experimental sample using our preferred doubly robust ATE estimator. Adjusting for confounder imbalance is theoretically unnecessary in randomized samples, but can improve stability in finite samples, as in \citet{wager2016}. We also trim the experimental sample to the region of propensity scores with common support in the \textit{observational} sample, using the $\hat{e}(\cdot)$ plotted in Figure \ref{fig:propensity_plots}, panel (c). This step gives us an experimental benchmark of the common support ATE, which is all we can estimate in the observational sample and which may differ from the full population ATE if treatment effects are heterogeneous. 

\subsection{Main Results: Following Best Practices}
\label{sec:bestpractices}

\begin{figure}[htbp]
    \centering
    \includegraphics[width=0.9\textwidth]{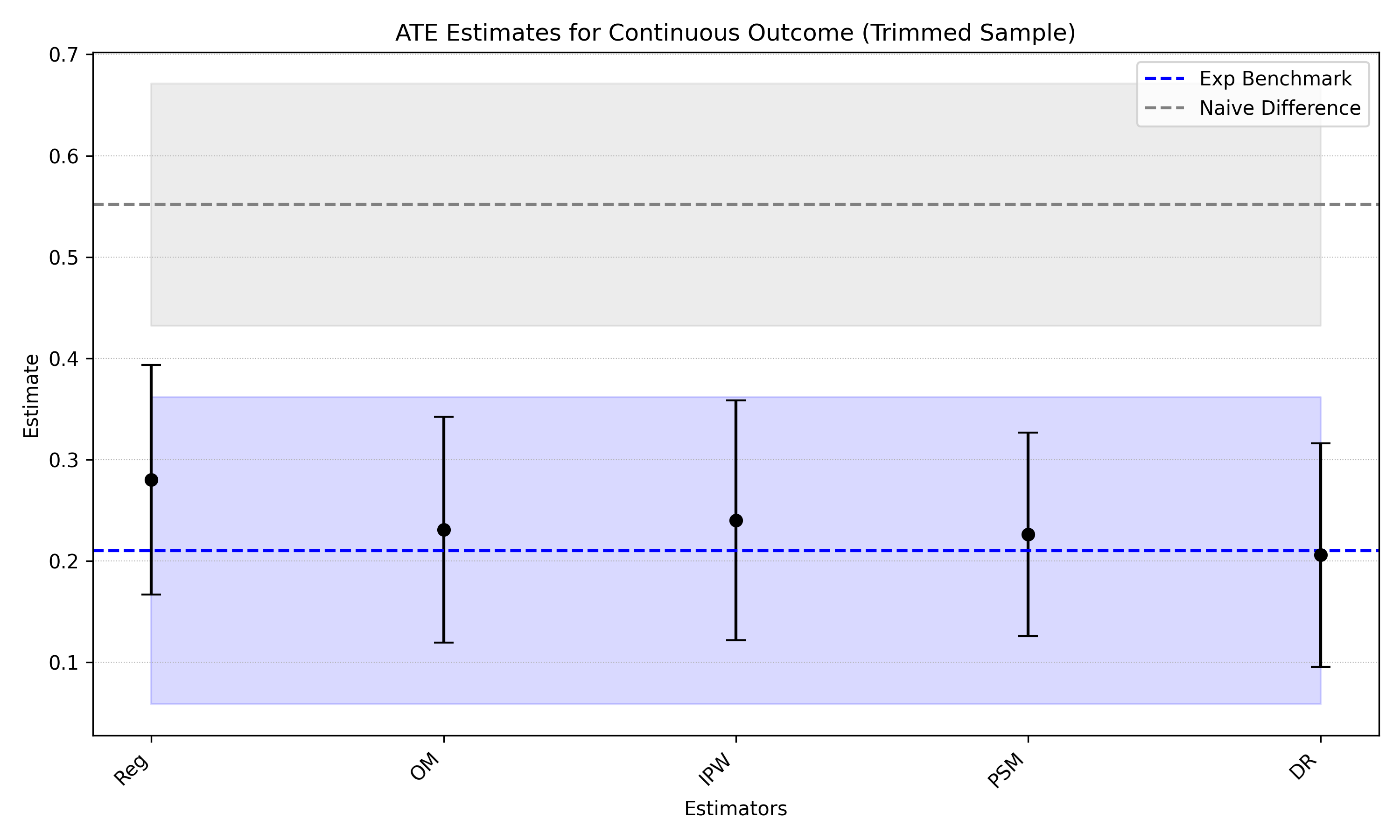}
    \caption{\textbf{ATE Estimates for General Performance Outcome (Trimmed Sample):} The figure displays ATE estimates and their 95\% confidence intervals for five estimators: regression adjustment (Reg), outcome modeling (OM), inverse probability weighting (IPW), propensity score matching (PSM), and doubly robust estimation (DR). After trimming, careful nuisance tuning, and model ensembling, all estimators align closely with the experimental benchmark.}
    \label{fig:ate_summary_trimmed}
\end{figure}

In the trimmed samples, we can recover the experimental baseline from the observed sample using most causal estimation methods introduced in the previous section. We estimate the average treatment effect (ATE) using the five estimators discussed in section \ref{sec:methodology}: linear regression (Reg), Oaxaca-Blinder outcome modeling (OM), inverse probability weighting (IPW), propensity score matching (PSM), and doubly robust AIPW estimation (DR).\footnote{In the linear regression framework, categorical features are one-hot encoded and continuous features enter linearly. This is the simplest version of a linear regression. While one could move linear regression toward the flexibility of tree-based estimators with the addition of interaction terms or flexible splines, but we think the ad hoc way these additions are often made in practice is less desirable than the hands-off approach of cross-fitting.} We plot these estimates against the experimental baseline in Figure \ref{fig:ate_summary_trimmed}. All five estimators include the ground truth within their 95\% confidence bands, though some hit closer with their point estimates than others. In this large data sample all five estimators also have similar confidence bands that approach the efficiency bound, although the LaLonde sample in the following section shows more variation.

In addition to sample trimming, the estimates in Figure \ref{fig:ate_summary_trimmed} reflect several other best practices for practical causal estimation: careful selection of the nuisance models and model ensembling. For the ``nuisance models" we use to construct our estimators, $\mu_d(X) \equiv \E[ Y \mid D=d, X]$ and $e(X)\equiv \E[D \mid X]$, we compare multiple models, including flexible machine learning models with hyperparameters tuned using AutoML procedures to maximize predictive performance, and choose the ones with the best fit. Finally, instead of relying on a single model for each nuisance, we ensemble across multiple models to mitigate model uncertainty. Following \citet{breiman1996bagging}, averaging across multiple learners reduces variance and stabilizes treatment effect estimation. 

The remainder of this section illustrates the costs of deviating from each of these best practices. While we find that many causal estimators can recover experimental benchmarks following all of our recommended practices, the following sections show the DR estimator is generally the most robust to deviations from these best practices. We therefore treat this estimator as our baseline when exploring extensions to our main results.

\subsection{Sample Trimming}
\label{sec:trimming}

Ensuring common support between treated and control units is critical for estimation. In settings with poor overlap, pruning the data to a region of overlap is often more important for credibility than the specific choice of estimation technique \citep{DehejiaWahba1999, DehejiaWahba2002}. The resulting estimand is the average treatment effect for the trimmed subpopulation rather than the full-sample ATE or ATT. We follow \citet{crump2009dealing} and arrive at an optimal trimming threshold of $[0.026,0.974]$.\footnote{In practice, given the substantial treatment imbalance in our setting, trimming is effectively one-sided.} As shown in Table \ref{tab:trimming_summary}, these cutoffs remove 32\% of the observational sample, consisting of 8\% of treated units and 33\% of control units. Implementing the same cutoffs in the experimental sample drops a similar share of total observations, as we would expect from Figure \ref{fig:propensity_plots} panel (c), but in this case balanced between treated and untreated units.

\begin{figure}[htbp]
    \centering
    \includegraphics[width=0.9\textwidth]{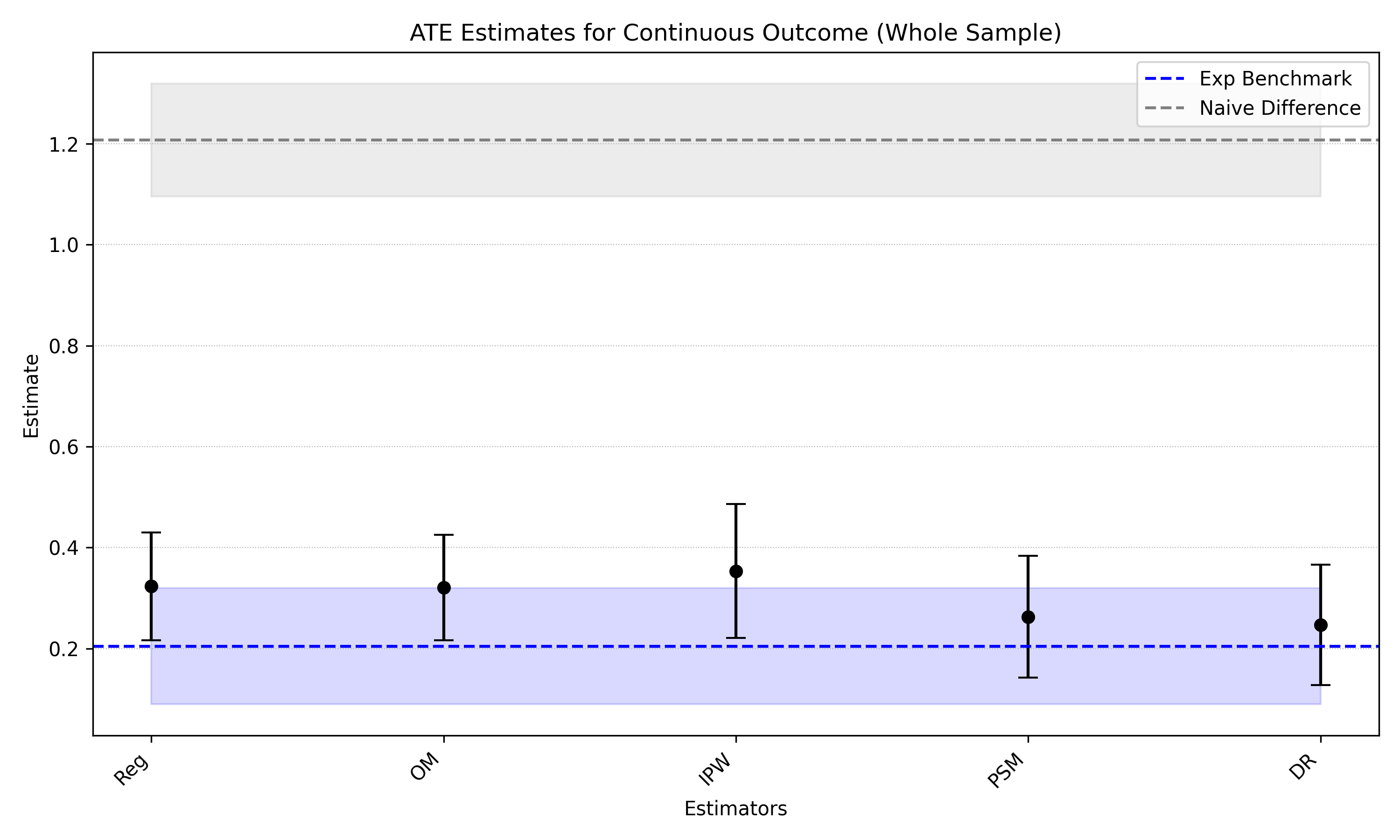}
    \caption{\textbf{ATE Estimates for General Performance Outcome (Untrimmed Sample):} The figure displays ATE estimates and their 95\% confidence intervals for five different estimators: regression adjustment (Reg), outcome modeling (OM), inverse probability weighting (IPW), propensity score matching (PSM), and doubly robust estimation (DR). The black dashed line and associated shaded band represent the naive difference in means and its 95\% confidence interval, while the blue dashed line and associated band show the experimental benchmark and its 95\% CI. Without trimming, estimates are more dispersed, with only DR encompassing the point estimate of the benchmark.}
    \label{fig:ate_summary_untrimmed}
\end{figure}

Figure~\ref{fig:ate_summary_untrimmed} reports ATE estimates and 95\% confidence intervals for the full, untrimmed sample, following our other best practices. The naive difference-in-means estimate is substantially farther from the experimental benchmark in the full sample. While all estimators close most of this gap, they are still biased somewhat upwards from the true treatment effect. Only the doubly robust estimator still contains the point estimate of the ground truth within its confidence band.

\subsection{Nuisance Model Tuning}
\label{sec:tuning}

Modern causal estimation methods like doubly robust learners outperform linear regression in our baseline comparisons because they use flexible machine learning models to better capture the relationships between confounders, treatment, and outcome. However, their performance is dependent on finding models that actually capture these relationships accurately. Flexibility can become a liability if the hyperparameters that control these models are chosen poorly and the models overfit to the training sample \citep{bach2024}. 

Table \ref{tab:first_stage_summary_combined} illustrates the out-of-sample performance of various models when estimating each of the ``nuisance" functions required for doubly robust estimation: the predicted outcome for the treated and untreated samples, $\hat{\mu}_0(X)$ and $\hat{\mu}_1(X)$, and the probability of treatment, $\hat{e}(X)$. The models are listed in descending order of their average performance across the three functions. Our preferred models are a well-tuned light gradient-boosted machine (LGBM) classifier to predict treatment, and light gradient-boosted machine regressors to predict the general outcome.\footnote{For LGBM and Lasso, we employ Microsoft's FLAML, a Bayesian optimization framework for hyperparameter tuning. We tune each nuisance function estimator for up to 100 seconds. For random forests and neural networks, we use Optuna since the current version of FLAML (as of this writing), has a limited hyperparameter search space for these models. Neural networks are never truly “untuned” in practice, and their weak performance on tabular data is well-documented. Our "weakly tuned" neural network models perform only a cursory hyperparameter search to illustrate this point; see Appendix Table \ref{tab:nn_tuning_spaces} for full details.} Using tuned LGBM, we can predict close to 13\% of the variation in the general outcome for treated and untreated units in the trimmed sample. These models also do the best job predicting the probability that each unit is treated, although this top performance still yields a modest AUC of just under 0.6. The performance of the propensity score model is consistently higher in the full sample. This pattern is unsurprising since the trimmed sample excludes groups of devices with very high or very low treatment shares, which are also the easiest to predict.

The remaining rows of Table \ref{tab:first_stage_summary_combined} illustrate a variety of alternative methods for estimating the nuisance functions.\footnote{The full propensity score distributions for these models are plotted in Appendix Figures~\ref{fig:Model_untrimmed_Comparison_PScore_Dist} and~\ref{fig:Model_trimmed_Comparison_PScore_Dist}. Without tuning, propensity score estimates vary widely across models, often failing to meaningfully separate treated and untreated groups.} In the software feature data, an untuned LGBM performs nearly as well as the tuned version, though we expect this is a coincidence of the default choices rather than a general feature of these models. For example, the default tree depth is faily shallow, which does well in our data sample but might be a bad fit for data with large categorical features like industry or country. Tuned random forest models also perform nearly as well as our baseline LGBM, though even a tuned neural net does not. Logistic/linear regression and penalized Lasso models perform notably less well, indicating that the linearity assumptions these models impose is not the best fit for our data. The performance of the tuned Lasso is nearly identical to the unpenalized logistic/linear regression models; for this relatively long and narrow data sample, the optimal Lasso penalty is close to zero. However, the bottom of Table \ref{tab:first_stage_summary_combined} is dominated by flexible models estimated at their default hyperparameter values. Without careful handling, these more complex models predict much less of the variation in both treatment and outcomes than a simple linear regression.
 
\begin{figure}[htbp]
    \centering
    \includegraphics[width=0.85\linewidth]{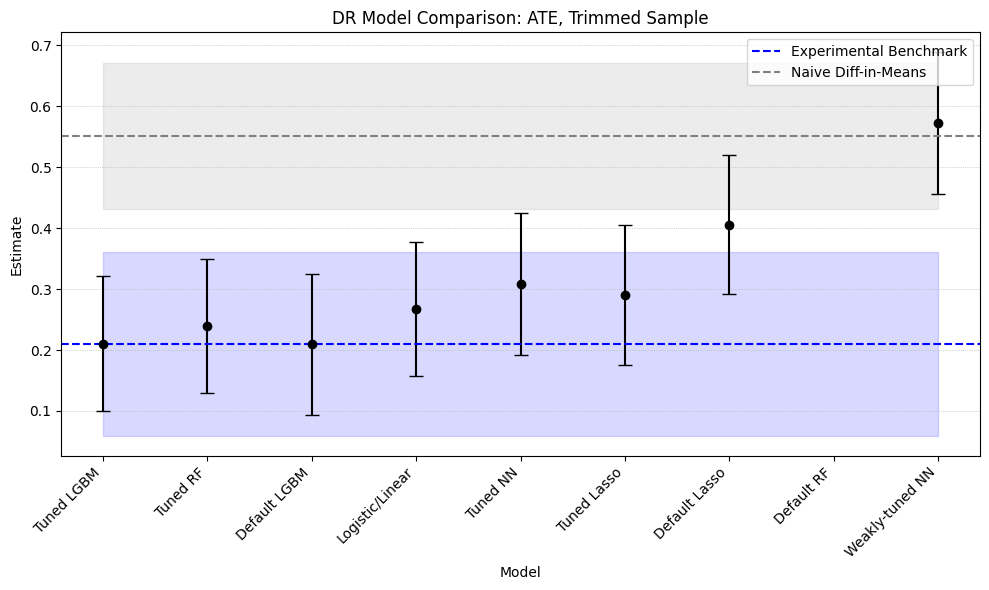}
    \caption{\textbf{DR Estimates Across Different First-Stage Models (Trimmed Sample):} This figure compares ATE estimates from doubly robust estimators using a range of flexible and rigid first-stage models for outcome and treatment modeling. Both tuned and sklearn default versions are considered. We find tuning is critical for flexible models, while its effect is less important with rigid models. Models like tuned LightGBM and tuned random forests align closely with the benchmark, while weakly tuned neural networks continue to exhibit bias. The default random forest implementation yields an estimate of -55.409 (83.877), driven by very unstable propensity weights, and is thus not pictured in the figure. In contrast to the untrimmed comparison in Figure \ref{fig:dr_comparison_untrimmed}, we see trimming substantively reduces estimator variability and brings many models closer to the experimental benchmark, partially mitigating the gains from tuning. These results altogether reinforce that trimming and tuning are complementary tools for improving the reliability of observational causal estimation.}
    \label{fig:dr_comparison_trimmed}
\end{figure}

To quantify the final impact of modeling choices, we re-estimate the doubly robust estimator across this range of first-stage nuisance models.  Figure~\ref{fig:dr_comparison_trimmed} plots the resulting ATE estimates, with the models again ordered by average performance. Across models, we observe a near-monotonic relationship between first-stage predictive quality and downstream causal estimate quality: models with higher $R^2$ and AUC scores produce estimates closer to the experimental benchmark. The downstream effects of nuisance model performance are quantitatively important; the worst-fitting models do no better than the unconditional difference-in-means estimator.

This exercise illustrates that choosing the right functional form for causal analysis is as quantitatively important as choosing the right set of conditioning variables. Happily, the best functional form is easy to identify. Modern methods that use machine learning to estimate underlying relationships make this step easier. Rather than ad hoc choices by the econometrician on what potential confounders to include and how to featurize them, ML methods identify complex, non-linear relationships automatically and allow practitioners to err on the side of inclusion\footnote{As long as the conditioning set only includes potential confounders and not ``bad controls" such as colliders or downstream outcomes. See \citet{pearl1995causal} or \citet{angrist2009mostly} for more discussion.} while avoiding overfitting through cross-estimation and penalization. For a given data sample, choosing functional forms that maximize predictive performance in the nuisance models is likely to minimize bias in the treatment effect estimates. We recommend trying multiple models, including tuned, flexible ML models, or using an AutoML algorithm to select across many models.

\subsection{Nuisance Model Stability}
\label{sec:ensemble}

Our preferred estimation approach involves two rounds of arbitrary sample splitting. We first use cross-validation to tune the hyperparameters for each nuisance function. Then, we cross-fit those nuisance functions within the doubly robust estimation to enable valid confidence interval construction in the final stage (see, i.e., \citet{cherono2018dml}). The previous section illustrates the importance of this approach for satisfying the unconfoundedness assumption. However, these cross-fitting steps also introduce a new, and often overlooked, source of uncertainty. For each iteration of cross-fitting or cross-validation, the data sample is randomly split into equal sized folds. Even when using well-tuned learners, different splits of the data can yield substantively different nuisance models \citep{ritzwoller2024reproducible}. Even if performance metrics for nuisance models are similar across runs, this variation can carry into downstream causal estimates. This \textit{model uncertainty} is distinct from \textit{sample uncertainty}, which arises because we estimate treatment effects on a finite sample of data drawn from a larger population of interest and is captured via standard errors. 

\begin{figure}[htbp]
    \centering
    \includegraphics[width=0.9\linewidth]{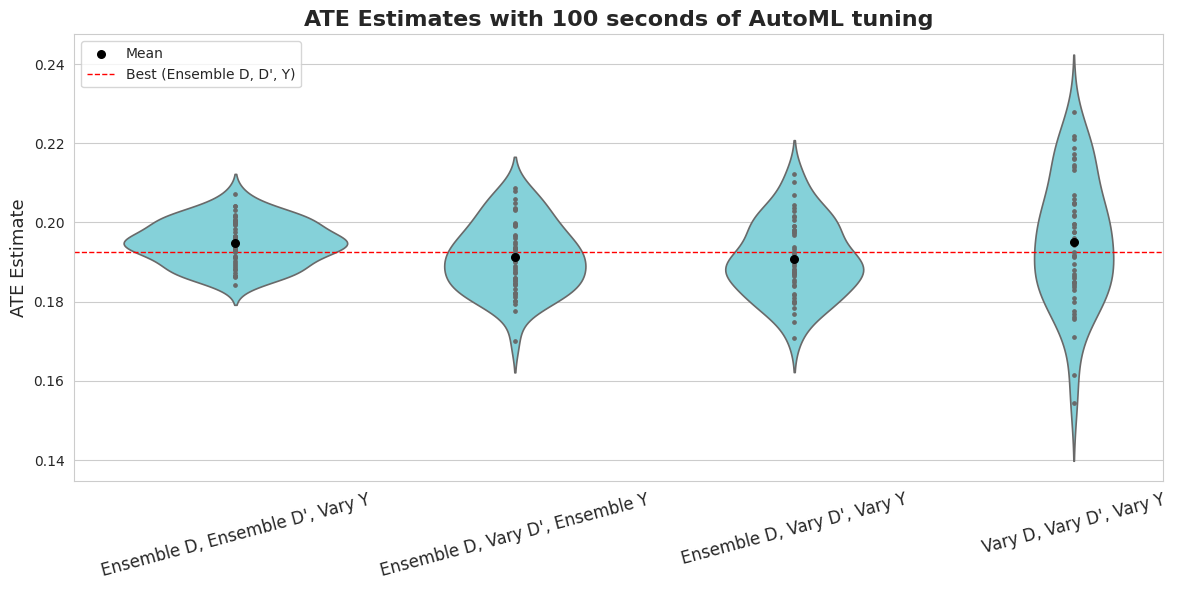}
    \caption{Model Uncertainty: ATE Estimates}
    \label{fig:ate_model_uncertainty}
    \caption*{\small \textit{Notes:} This figure illustrates model uncertainty in ATE estimates arising from variability in nuisance function estimation. The fully ensembled learner is shown by the red dotted line. Starting from the whole sample, AutoML is run a total of 150 times: 50 trials for the initial propensity model $\hat{e}(X)$, and 50 trials each for the re-estimated propensity model $\hat{e}'(X)$ and the outcome models $\hat{\mu}_0(X)$ and $\hat{\mu}_1(X)$ after trimming, with different random seeds affecting cross-validation splits during hyperparameter search. Fixed models are then combined according to four stabilization strategies: (i) ensemble $\hat{e}(X)$ and $\hat{e}'(X)$, while varying $\hat{\mu}_0(X)$ and $\hat{\mu}_1(X)$; (ii) ensemble $\hat{e}(X)$, while varying both $\hat{e}'(X)$,  $\hat{\mu}_0(X)$, and $\hat{\mu}_1(X)$; (iii) ensemble $\hat{e}(X)$,  $\hat{\mu}_0(X)$ and $\hat{\mu}_1(X)$, while varying  $\hat{e}'(X)$; and (iv) vary all three nuisances without ensembling any. Results show that ensembling even a subset of nuisance models reduces dispersion in ATE estimates, highlighting the benefits of such a procedure to stabilize flexible nuisance learners.}
\end{figure}

In our recommended approach, we ensemble 50 tuned LGBM models for each of four preliminary models: the initial propensity model used to trim the sample, which we denote $e^*(X)$, the propensity model re-estimated on the trimmed sample within the doubly robust algorithm, $e(X)$, and the two outcome models, $\hat{\mu}_0(X)$ and $\hat{\mu}_1(X)$. As discussed in \citet{breiman1996bagging}, averaging across models improves predictive accuracy and mitigates downstream instability. In Figure \ref{fig:ate_model_uncertainty}, the point estimate for this fully ensembled baseline is marked by the red dotted line. The remainder of the figure illustrates the potential instability created by foregoing this ensembling. The first shape plots the range of treatment effects estimated using each of the 50 single runs of the outcome models while continuing to average across runs for the two propensity models. The remaining shapes conduct similar exercises, plotting the distribution when:

\begin{itemize}
    \item Ensembling $\hat{e}^*(X)$, $\hat{\mu}_0(X)$, and $\hat{\mu}_1(X)$, while varying $\hat{e}(X)$ over 50 trials.
    \item Ensembling $\hat{e}^*(X)$ and varying  $\hat{e}(X)$, $\hat{\mu}_0(X)$, and $\hat{\mu}_1(X)$ over 50 trials.
    \item Varying all  components over 50 trials.
\end{itemize}

Figure \ref{fig:ate_model_uncertainty} illustrates the real variation that can occur between runs of these models with different seeds, particularly the last shape where none of the models are ensembled, a common starting point in empirical work. Comparing this figure with our baseline estimates in Figure \ref{fig:ate_summary_trimmed} illustrates that the instability introduced by random sample splitting is considerably smaller than our standard errors, which capture the uncertainty created by drawing the initial data sample from the population. This is also a smaller source of instability than the model selection discussed in the previous section. Ensembling across multiple runs of the nuisance functions can add considerably to run times, particularly when each run includes re-tuning hyperparameters, as we believe it should. For high-stakes estimates where precision matters, this extra time can be a worthwhile investment.

\section{Additional Samples}\label{sec:addl_samples}

In the previous section, we illustrate the importance of three key elements of our recommended best practices using the estimated effect of a new product feature on one general outcome in our data sample. We now briefly explore how robust those recommendations are across different outcome types and datasets

\subsection{Impact of New Feature on Targeted Outcome}

\begin{figure}[htbp]
    \centering
        \includegraphics[width=\textwidth]{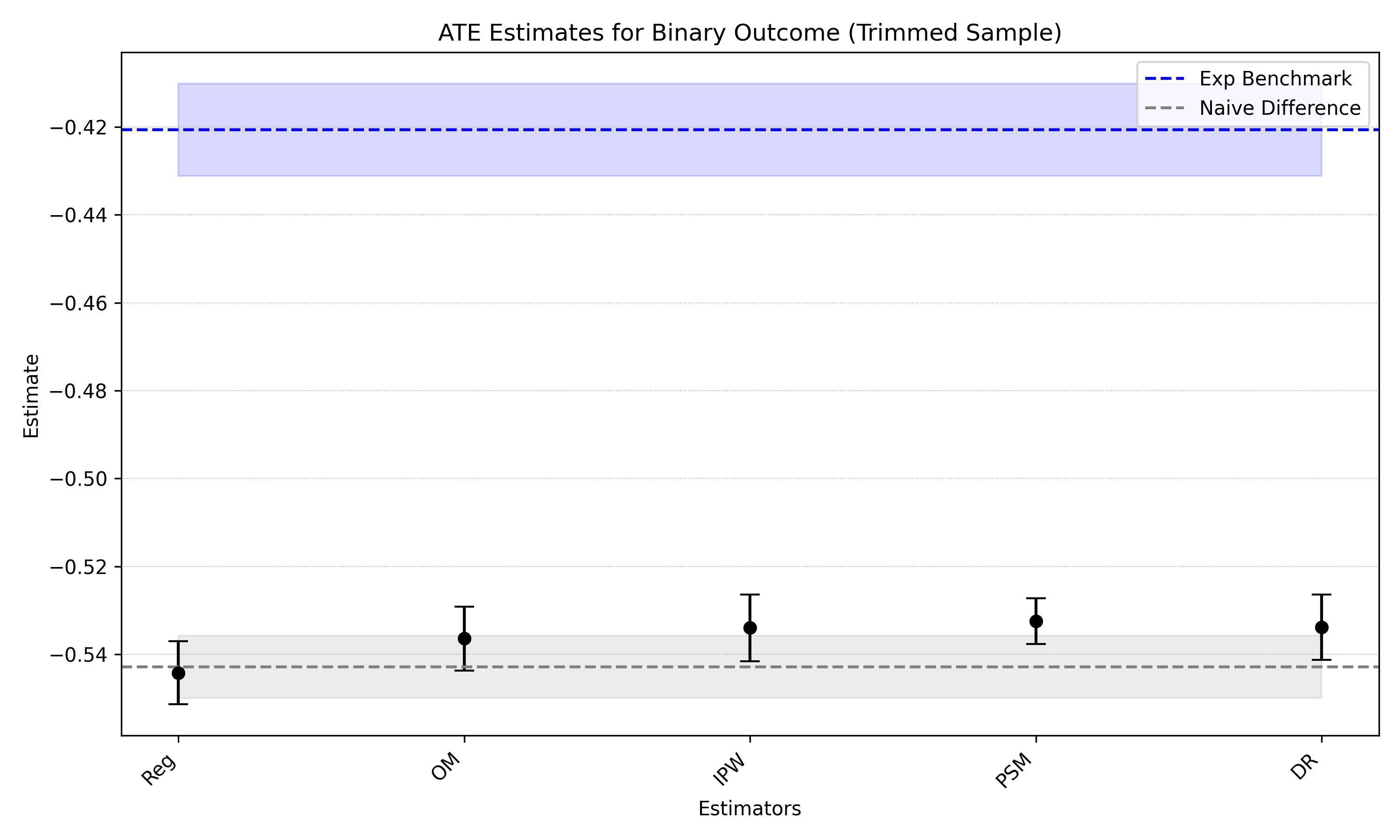}
    \caption{\textbf{ATE Estimates for Targeted Outcome:} ATE estimates and their 95\% confidence intervals for five estimators: regression adjustment (Reg), outcome modeling (OM), inverse probability weighting (IPW), propensity score matching (PSM), and doubly robust estimation (DR) in the trimmed software feature release sample. The blue dashed line and shaded band denote the experimental benchmark and its 95\% CI, respectively.}
    \label{fig:ate_summary_binary}
\end{figure}

We first consider the impact of the same new product feature on the binary outcome that was the primary target of the new feature. We use the same trimmed sample as in the general performance analysis, since trimming depends only on estimated treatment propensity and is independent of the outcome. However, as illustrated in Figure \ref{fig:ate_summary_binary}, for this outcome even the best practices miss the experimental benchmark by a wide margin; in fact, they are barely distinguishable from the naive difference-in-means.

This systemic pattern of bias across many estimates of the treatment effect suggests the presence of one or more unobserved confounders. Consistent with this hypothesis, we do a poor job of predicting the targeted outcome with our set of available measured confounders. The AUC for our preferred $\hat{y}_1$ model is 0.54 in the trimmed observational sample, barely better than random guesses.\footnote{The AUC to predict the targeted outcome for untreated units is higher, 0.82, but it is difficult to measure performance for this group as nearly 99\% of units have a positive outcome (see Appendix Table \ref{tab:within_sample_smd}).} A missing confounder also seems plausible for this particular data sample. The product group that collected both the experimental and observational data samples included a set of standard device and user characteristics that they use as conditioning variables in many analyses. It seems likely that this set of characteristics would do a good job capturing endogenous drivers of the general performance outcome, but miss some important, idiosyncratic drivers of the specific performance metric targeted by the new feature.

This finding illustrates an important reminder: well-fit, flexible ML models make it easier to achieve unconfoundedness by extracting all the available signal from a set of conditioning variables. However, even the best methods cannot overcome important omissions from this set.

\subsection{Re-analyzing LaLonde}

\citet{xuimbens2025} deploy most of our suggested best practices to re-analyze the worker training samples introduced in \citet{LaLonde1986}. For comparison, we briefly repeat the exercise here. We follow the convention in \citet{DehejiaWahba1999}, referring to the reanalyzed LaLonde samples as LDW-CPS and LDW-PSID, which pair experimental treated units with controls from the CPS and PSID, respectively. We apply four estimators (doubly robust, propensity score matching, inverse propensity weighting, and outcome modeling) using \cite{crump2009dealing} trimming, AutoML tuned nuisances, and model ensembling. To ease comparison with much of the long line of earlier studies using these data samples, we estimate the average treatment effect on the treated (ATT) for these samples, rather than the population ATE we consider for the software feature release sample.

\begin{figure}[htbp]
    \centering
    \begin{subfigure}[t]{0.85\textwidth}
        \centering
        \includegraphics[width=\textwidth]{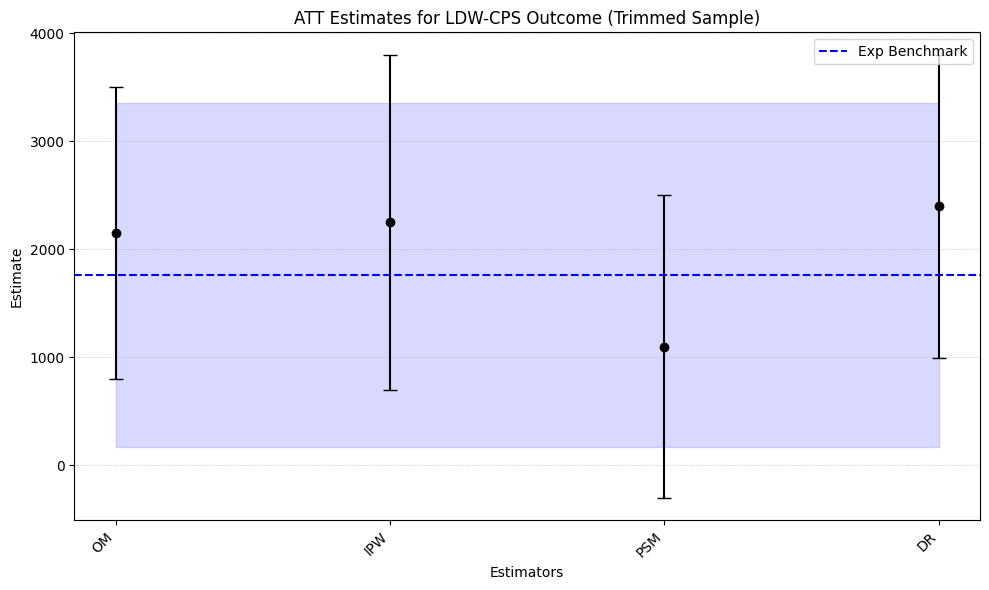}
        \caption{LDW-CPS Sample}
    \end{subfigure}
    \vspace{0.5em}
    \begin{subfigure}[t]{0.85\textwidth}
       \centering
        \includegraphics[width=\textwidth]{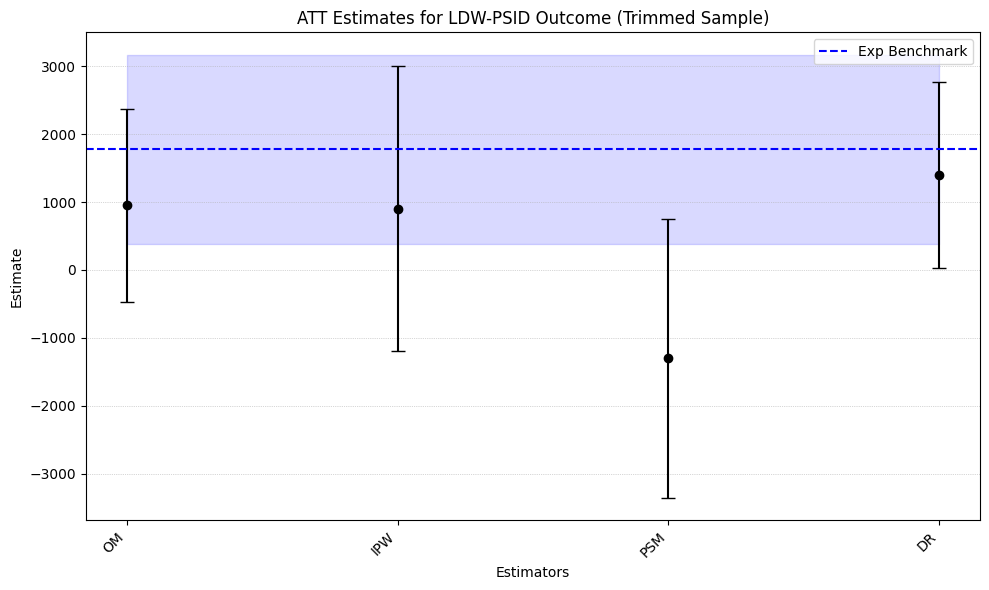}
       \caption{LDW-PSID Sample}
   \end{subfigure}
    \caption{ATT Estimates: Each point represents an estimator’s ATT estimate with 95\% confidence intervals. The blue dashed line and shaded band denote the experimental benchmark and its 95\% CI, respectively.}
    \label{fig:lalonde_summary}
\end{figure}

Figure \ref{fig:lalonde_summary} illustrates the results of this exercise, which confirm the findings of Xu and Imbens. In the trimmed LDW-CPS sample, all four estimators yield ATT estimates between approximately \$1,100 and \$2,400, and all fall within the experimental benchmark confidence band (centered at \$1,761). The LDW-PSID results display greater heterogeneity. The PSM estimator is highly negative (near -\$1,300), consistent with the instability reported in earlier studies. IPW, OM, and DR estimators center around \$1,000, well below the experimental benchmark of \$1,775, but have wide confidence intervals that contain the benchmark. 

As discussed in more detail in \citet{DehejiaWahba1999} and many others, the lack of overlap in these LaLonde samples is severe. Our main departure from \citet{xuimbens2025} is in how we trim the samples to create a cleaner comparison. Xu and Imbens trim the treated workers based on rule-of-thumb thresholds, $[0.1,0.9]$ for the LDW-CPS sample and $[0.2,0.8]$ for LDW-PSID, and then select a 1:1 matched set of control observations. The \citet{crump2009dealing} procedure we follow selects propensity score thresholds of 0.085 and 0.915 for the LDW-CPS sample and 0.109 and 0.891 for the LDW-PSID sample. These thresholds allow us to retain somewhat more of the sample, particularly for the LDW-PSID sample, while matching or slightly improving the performance of most estimators.

\section{Extensions and Caveats}

We close by extending our main analysis along several dimensions. 
First, assess the ability of sensitivity analysis and other tests to detect when the unconfoundedness assumption required by all our methods is likely to be violated. Next, we explore heterogeneous treatment effect estimation, assessing whether modern methods can reliably recover patterns of treatment effect variation across covariates.

\subsection{Sensitivity Analysis} 
\label{sec:sensitivity}

In the previous sections we explored the effect of a new software feature on two outcomes. For the general device performance metric, we are able to recover experimental benchmark effects using careful causal estimation methods. For the targeted performance outcome, no method recovers the experimental benchmarks. We hypothesize that this failure is consistent with an unobserved feature that confounds only the relationship between the new software feature and the targeted performance outcome. There are no direct tests for the presence of such a confounder. Sensitivity analyses \citep{cinelli2019, chernozhukov2022} can illustrate how robust our estimates are to a hypothetical unobserved confounder.

Our sensitivity analysis follows \citet{chernozhukov2022} and quantifies how strongly an omitted confounder $U$ would need to be related to the outcome $Y$ and treatment $D$ in order to overturn our causal conclusions. Three key quantities characterize this relationship:

\begin{itemize}
    \item {$C_Y^2$}. The nonparametric partial $R^2$ of an omitted confounder $U$ with the outcome $Y$, conditional on observed covariates $X$ and treatment $D$:
    \[
    C_Y^2 \;=\; 
    \frac{\text{Var}(\mathbb{E}[Y|D,X,U]) - \text{Var}(\mathbb{E}[Y|D,X])}{\text{Var}(Y) - \text{Var}(\mathbb{E}[Y|D,X])}
    \]
    This measures the additional share of unexplained variation in $Y$ that could be explained by $U$ once $X$ and $D$ are controlled for.

    \item {$C_D^2$}. The relative gain in precision in predicting treatment assignment when $U$ is added to the observed covariates:
    \[
    C_D^2 \;=\;
    \frac{\mathbb{E}\left[\left(P(D=1|X,U)(1-P(D=1|X,U))\right)^{-1}\right] - \mathbb{E}\left[\left(P(D=1|X)(1-P(D=1|X))\right)^{-1}\right]}{\mathbb{E}\left[\left(P(D=1|X,U)(1-P(D=1|X,U))\right)^{-1}\right]}
    \]
    This captures how much sharper the propensity score becomes once $U$ is observed.

    \item {$\rho$}. The correlation between the omitted confounder’s effect on $Y$ and its effect on $D$, conditional on $X$:
    Values of $|\rho|$ close to 1 correspond to a confounder that moves both outcome and treatment in the same direction, generating maximum bias.
\end{itemize}

These quantities bound the potential bias between the feasible estimate, $\tau$, and the unconfounded estimate that conditions on $U$, $\tau_S$:
\[
|\tau_S - \tau| \;\le\; |\rho|\,C_Y\,C_D\,S,
\]
where $S$ is a sensitivity scaling factor. Intuitively, $S$ rescales the bound by the variability of the outcome residuals and the informativeness of the treatment assignment, so that $C_Y$ and $C_D$ are measured on a unit-free scale. Full details on its empirical construction are provided in Appendix~\ref{app:sensitivity-implementation}.

Given $(C_Y, C_D, \rho)$, the maximum bias in the ATE estimate can be bounded. Two summary metrics are of particular interest. First, we calculate the robustness value ($RV(\rho)$), the minimum explanatory strength of an unobserved confounder, that is a minimum value of $C_Y = C_D$, required to shift the true point estimate of the treatment effect to zero. Second, we report a weaker adjusted robustness value ($\text{RV}_\alpha(\rho)$), the minimum explanatory strength required to shift the confidence interval of the estimate to include zero. For our general performance ATE estimate, assuming a scenario where $\rho=0.5$, we obtain: {RV = 1.31\%} and {$\text{RV}_\alpha(\rho)$ = 0.60\%}.\footnote{While $\rho=0.5$ is not the worst-case confounding scenario, it provides a conservative value relative to the $\rho$ we estimate for any of our observed confounders, reported in Appendix Table \ref{tab:covariate-benchmarking}. Among the observed confounders, all variables with meaningful influence on the treatment or outcome have a $\rho$ well below 0.5.} An omitted confounder would need to explain 1.31\% of the residual variation in both $Y$ and $D$ to drive the point estimate to zero, or 0.6\% of the residual variation in both $Y$ and $D$ to move the confidence interval to include zero.

These robustness values are small in absolute terms, but they are difficult to interpret without benchmarks of plausible sets of $\left\lbrace C_Y, C_D, \rho \right\rbrace$. We can calculate these values for each observed covariate in our sample by excluding them one-by-one and re-estimating the model. For the general outcome, the strongest overall confounder is the total device usage. For this feature, $C_Y=0.023$, $C_D = 0.013$, and $\rho=0.104$.\footnote{Values for all observed confounders are reported in Appendix Table \ref{tab:covariate-benchmarking}. The indicator that Engagement Segment = Highly Engaged has a stronger role in predicting treatment, $C_D=0.015$, but far less influence on the outcome, $C_Y=0.00038$, with $\rho = 0.207$.} These influence values are well above the robustness value, suggesting that the presence of a similarly influencial unobserved confounder is plausible, although we emphasize that the $\rho$ we estimate for this feature is well below 0.5. An unobserved confounder with $\rho=0.1$ would generate far less bias.

The robustness values we calculate for the targeted outcome are considerably higher. To move the estimate to zero, an omitted confounder would need to explain approximately 74.8\% of the residual variation in both the treatment and outcome.
Identifying an appropriate benchmark for these values in the targeted outcome is more difficult. As we have already established, none of our observed confounders do a good job predicting the targeted outcome; the highest $C_Y$ we estimate is 0.0026 for the indicator Manufacturer=5. Instead of comparing a hypothetical unobserved confounder to this set of weak predictors of the targeted outcome, we think a more reasonable benchmark is to use the same influence values as the strongest observed confounder for the general outcome.

\begin{figure}[htbp]
    \centering
    \begin{subfigure}[t]{0.48\textwidth}
        \centering
        \includegraphics[width=\textwidth]{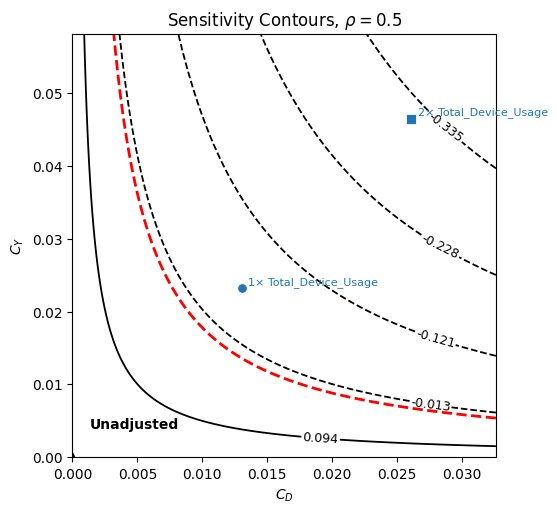}
        \caption{General performance outcome}
    \end{subfigure}
    \hfill
    \begin{subfigure}[t]{0.48\textwidth}
        \centering
        \includegraphics[width=\textwidth]{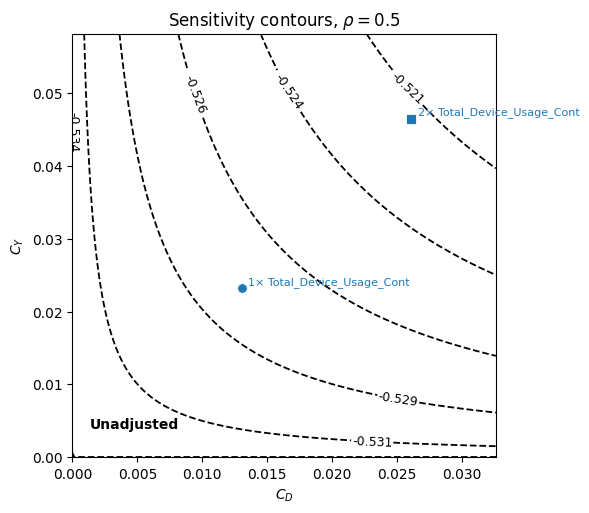}
        \caption{Targeted outcome}
    \end{subfigure}
    \caption{\textbf{Sensitivity Analysis Results:} Panel (a) displays bias contours for the general performance outcome, assuming correlation $\rho=0.5$ between unobserved confounding effects on treatment and outcome. Moving north-east, each dashed line indicates greater estimated bias relative to the original point estimate. The ``unadjusted'' is our estimate without unobserved confounding. The dashed red line represents a zero effect. The dots plot the treatment effect we would estimate if we added confounders with 1x and 2x the ($C_Y$, $C_D$) as $\mathbb{I}\{\text{Total\_Device\_Usage}\}$ 
    at $\rho=0.5$, which is a conservative estimate given what we observe in our data, shown in our benchmarking exercise in Appendix Table \ref{tab:covariate-benchmarking}. Panel (b) shows the same hypothetical confounder but in the targeted outcome setting.}
    \label{fig:sens_contour}
\end{figure}

Figure \ref{fig:sens_contour} plots a range of impacts on our treatment effect estimates for each outcome of various unobserved confounders. The x-axis shows values for $C_D$, the y-axis values of $C_Y$ and the isoquants plot the treatment effect we would have estimated if we were able to include an additional confounder of various strengths in our analysis. The ``unadjusted" dot at the origin marks our baseline treatment effect estimate. The blue dots plot the treatment effect we would estimate if we added confounders with 1x and 2x the benchmark ($C_Y$, $C_D$)  of $\mathbb{I}\{\text{Total\_Device\_Usage}\}$ for the general outcome. These plots reinforce the message of the robustness values. For the general outcome, plotted in the left panel, a plausible unobserved confounder could easily push our estimated treatment effect past zero (indicated with a dashed red line). In contrast, for the targeted outcome in the right panel, a benchmark unobserved confounder would have only modest effects, moving the estimated treatment effect from just below -0.53 to just below -0.52.

These contrasting results seem puzzling at first. Sensitivity analyses suggest that our estimated treatment effects on the general outcome, which we have demonstrated achieve the experimental benchmark, are highly sensitive to a hypothetical unobserved confounder. In contrast, our estimated treatment effects for the targeted outcome, which exhibit considerable bias relative to the experimental benchmark, are identified as largely robust to a hypothetical additional confounder. The solution lies in details of these sensitivity tests and the magnitude of the treatment effects.  

Comparing the ground truth treatment effects to the summary statistics in Appendix Table \ref{tab:within_sample_smd} indicates that our treatment increases the general outcome by about 0.03 standard deviations. The baseline ATE is 0.021, while the standard deviation of the general outcome is 6.7\footnote{We use the standard deviation for untreated observations in the experimental sample for this exercise.}. While statistically distinguishable from zero, this is a very small effect. The modest treatment effect also means that even the treatment and the other confounders combined explain relatively little of the outcome.\footnote{From Table \ref{tab:first_stage_summary_combined}, the confounders explain 13\% of the variation in the outcome.} An unobserved confounder would not need to explain much of the residual variation in order to push the very small treatment effect to zero.

In contrast, our ground truth estimates from the experimental sample indicate that our treatment decreases the targeted outcome by 3.6 standard deviations. Our best estimate from the observational sample suggests a decrease of 4.5 standard deviations, overstating the true effect but telling the same qualitative story. These are quantitatively important treatment effects, and the treatment drives much of the variation in the targeted outcome. In the experimental sample, predicting the targeted outcome using only the treatment yields an AUC of 0.86. Once we condition on both the treatment and the observed confounders, there is much less residual variation in the targeted outcome. An unobserved confounder would need to explain quite a lot of it to meaningfully change our estimates.

This example highlights both the value and the limitations of this type of sensitivity analysis. In this special case, we know that the estimated treatment effect on general performance, while small, represents a robust relationship that aligns with the experimental benchmark. In the common case, however, it is worth realizing that this very small treatment effect, estimated in a setting where much of the variation in the outcome remains unexplained, could easily be sensitive to even small missing confounders. In this special case, we know that our estimated treatment effects on the targeted outcome are somewhat biased and miss the experimental benchmark. However, both the experiment and the observational analysis capture the right qualitative message: the treatment has a strong and important negative effect on this outcome. The sensitivity analysis correctly identifies that adding additional confounders is unlikely to change that result. In some cases, as in the LaLonde samples analyzed in \citet{xuimbens2025}, the estimates identified as most vulnerable using a sensitivity check are also the estimates that are most biased away from the experimental ground truth. This exercise is a useful reminder that this alignment is not guaranteed. 

\subsection{Heterogeneous Treatment Effects}

Thus far we have evaluated the ability of causal inference methods to recover the average effect of treatment across the entire population of interest. However, many modern applications of causal analysis also require estimating treatment effects within various subgroups, or at least testing for the presence of treatment effect heterogeneity. The software feature release sample has the potential to be a good testing ground for methods that estimate these Conditional Average Treatment Effects (CATE). The large sample size should provide enough statistical power to estimate nuanced heterogeneous effects. In the experimental sample, randomized assignment ensures a thick distribution of treated and untreated observations across the full range of covariates, providing tight benchmark CATE estimates for all combinations of features. Modern tools offer a variety of flexible meta-learners for estimating CATE, including the S-, T-, X-, DR-, and R-learners. We first estimate each of these models separately to see if any fit the data significantly better than a constant treatment effects model, then combine them to compare the results of this best-practice CATE estimator against the experimental benchmark.

We evaluate the presence of treatment effect heterogeneity for each outcome in the software feature release data. The DR-score \citep{chernozhukov2024applied} measures the out-of-sample reduction in residual variance of the doubly robust pseudo-outcome $Y^{DR}$ relative to a constant CATE model. We begin by constructing the doubly robust variables, 
\begin{align}
Y_i^{DR}(\hat{\mu},\hat{e}) := \hat{\mu}_1(X_i) - \hat{\mu}_0(X_i) + (Y_i - \hat{\mu}_{D_i}(X_i))\frac{D_i - \hat{e}(X_i)}{\hat{e}(X_i) (1-\hat{e}(X_i))}
\end{align}
and then we train different CATE models $\hat{\tau}$ by regressing $Y_i^{DR}(\hat{\mu},\hat{e}) \sim X_i$. For any CATE model, we define the (normalized) DR-score: 
\begin{align}
\text{DRscore}(\hat{\tau}) := \frac{E_n\left[\left(Y^{DR}(\mu,e) - \tau_{\text{const}}(X)\right)^2\right] - E_n\left[\left(Y^{DR}(\mu,e) - \hat{\tau}(X)\right)^2\right]}{E_n\left[\left(Y^{DR}(\mu,e) - \tau_{\text{const}}(X)\right)^2\right]}
\end{align}

\noindent A score of 0 corresponds to a constant CATE model that predicts the ATE, and a score of 1 would reflect perfect prediction of $Y^{DR}$. Any CATE model that captures true heterogeneity should achieve a positive score between 0 and 1. 

\begin{figure}[htbp]
\begin{subfigure}{.475\linewidth}
  \includegraphics[width=\linewidth]{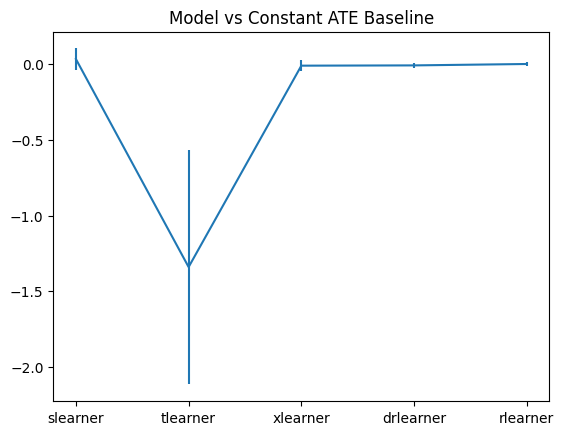}
  \caption{General performance outcome, experimental dataset}
  \label{sfig:cate-score-cont-exp}
\end{subfigure}\hfill
~ 
\begin{subfigure}{.475\linewidth}
  \includegraphics[width=\linewidth]{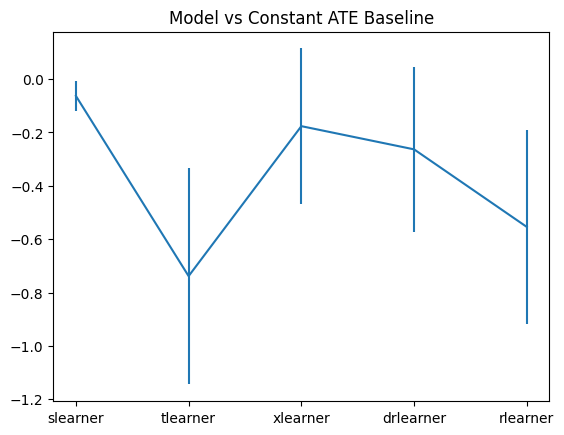}
  \caption{General performance outcome, observational dataset}
  \label{sfig:cate-score-cont-obs}
\end{subfigure}
\medskip
\begin{subfigure}{.475\linewidth}
  \includegraphics[width=\linewidth]{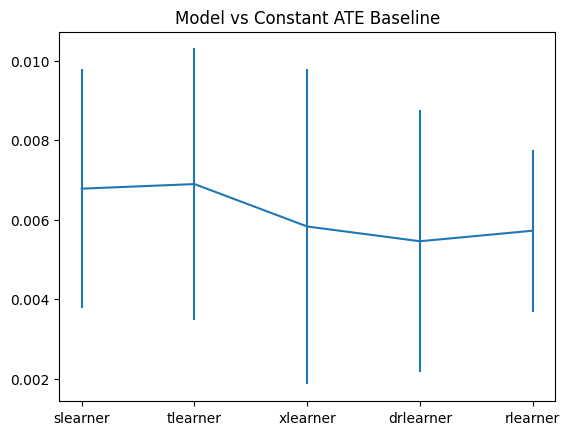}
  \caption{Targeted outcome, experimental dataset}
  \label{sfig:cate-score-binary-exp}
\end{subfigure}\hfill
~
\begin{subfigure}{.475\linewidth}
  \includegraphics[width=\linewidth]{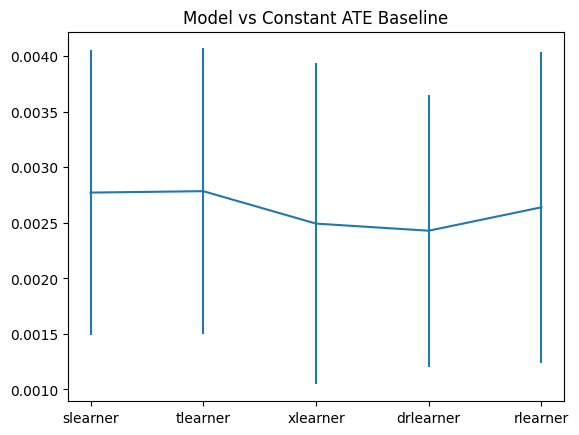}
  \caption{Targeted outcome, observational dataset}
  \label{sfig:cate-score-binary-obs}
\end{subfigure}
\caption{DR scoring for presence of heterogeneous effects in the software feature release data sample}
\label{fig:cate-scores}
\end{figure}

Figure \ref{fig:cate-scores} displays the DR-scores for each learner for our two outcomes, separately for the experimental and observational samples. Beginning with the experimental (left) column, we see that the DR scores are negative or zero across all models for the general performance outcome, indicating that none of the CATE models can outperform a constant effects model, and in fact estimating heterogeneous effects appears to add noise. For the targeted outcome, all CATE models modestly outperform the constant effects baseline, suggesting some heterogeneous effects. Reassuringly, conducting the same test in the observational sample (right column) yields the same results: the DR scores reject treatment heterogeneity for the general performance outcome and shows evidence of modest heterogeneity for the targeted outcome. 

While Figure \ref{fig:cate-scores} provides encouraging evidence on the ability of DR-scores to detect treatment heterogeneity in practice, it tells a discouraging story for the value of the software feature release data as a testing ground for CATE estimators. We only find evidence of variation in treatment effects on the targeted outcome. However, as shown in Section \ref{sec:addl_samples}, we cannot predict this outcome reliably from the available observed covariates and the ATE estimates in the observational sample show substantial bias relative to the experimental baseline, suggesting the presence of unobserved confounders. We proceed with an investigation of how well CATE methods used in the observational sample can replicate the \textit{ordering} of treatment effects on the targeted outcome across subgroups, with the caveat that we expect that all subgroups may show the same bias as the ATE estimates.

Our best-practice CATE estimates combine the models plotted in Figure \ref{fig:cate-scores}. Each of these methods exhibits its own bias-variance tradeoff and susceptibility to overfitting. Q-aggregation addresses this challenge by forming a convex combination of learners that penalizes individual learners that perform poorly \citep{lan2024causal}. Q-aggregation selects a vector of weights, $w$, that solves the penalized optimization problem:
\begin{equation}
\min_{w \in \Delta(M)} (1 - \nu) \mathbb{E}_n\left[ \left(Y^{DR} - \bar{\tau} - \sum_{m=1}^M w_m \tilde{\tau}_m(Z)\right)^2 \right] + \nu \sum_{m=1}^M w_m \mathbb{E}_n\left[ \left(Y^{DR} - \bar{\tau} - \tilde{\tau}_m(Z)\right)^2 \right],
\end{equation}
where $\tilde{\tau}_m(Z)$ denotes the centered CATE predictions from each model $m$, and $\Delta(M)$ is the simplex over $M$ models. The parameter $\nu \in [0,1]$ interpolates between convex regression, or OLS regression subject to simplex constraints ($\nu = 0$) and simply selecting the best model ($\nu = 1$).

The key strength of Q-aggregation lies in its ability to penalize learners that overfit to training noise or exhibit poor out-of-sample generalization. Unlike naive averaging, which may dilute strong learners, or hard selection, which risks high variance, Q-aggregation strikes a balance, selecting a convex combination that minimizes doubly robust loss. This procedure is supported by oracle inequality guarantees \citep{lecue2014optimal}, and in practice, it has proven effective for stabilizing CATE performance. Q-aggregation serves as both a model selection tool and a heterogeneity regularizer by discarding models that introduce spurious heterogeneity.

\begin{figure}[htbp]
\begin{subfigure}{.475\linewidth}
        \includegraphics[width=\linewidth]{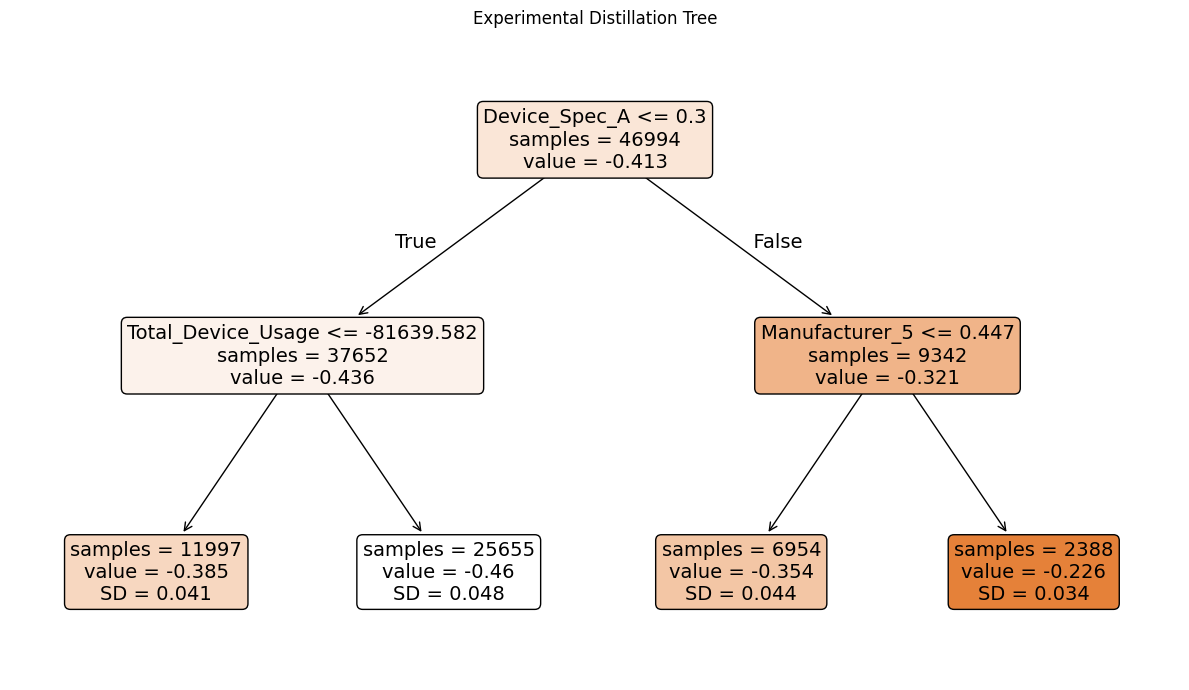}
        \caption{Experimental Sample}
\end{subfigure}\hfill
~
\begin{subfigure}{.475\linewidth}
        \includegraphics[width=\linewidth]{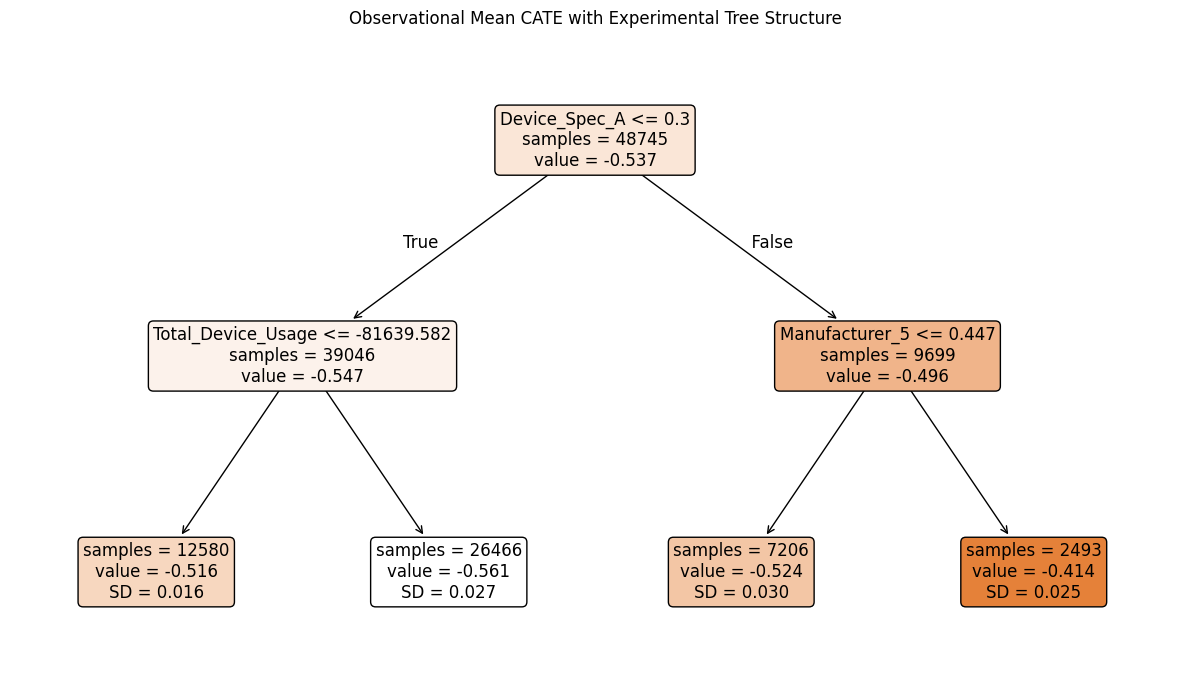}
        \caption{Observational Sample}
\end{subfigure}\hfill
    \caption{Distilled CATE trees from Q-aggregated models for the targeted outcome. In the experimental sample (left), the same variable splits are detected to minimize overlap across leaves. In the observational sample (right), we estimate CATE within the splits detected in the experimental sample. Each final leaf shows the standard deviation of estimated CATE within the group (not the standard error of the estimate).}
    \label{fig:distilled_cate_trees_binary}
\end{figure}

To create a benchmark for CATE estimation, we first use a shallow tree to distill what the Q-aggregated CATE ensemble models have learned in the experimental sample about the treatment effect on the targeted outcome. As shown in the left panel of Figure \ref{fig:distilled_cate_trees_binary}, the tree splits first on Device Specification A, then on Total Device Usage and an indicator of whether the device was made by Manufacturer 5. These cuts split the experimental sample into groups with substantially different average treatment effects. The group with the most negative treatment effect has an average CATE of -0.46 (sd = 0.048). The group with the smallest effect in absolute terms has an average CATE of -0.226 (sd = 0.034). 

We then split the observational sample into groups using these same feature values and consider the average CATE in each group as estimated from the observational sample. As shown in the right panel of Figure \ref{fig:distilled_cate_trees_binary}, the Q-aggregated CATE ensemble model is able to recreate the ordering of effect sizes across groups: the cluster with the most negative treatment effects in the experimental sample (Device Specification A is false, with high Total Device Usage) also has the most negative estimated effects in the observational sample and so on. However, as anticipated, all groups show the same systemic bias as the ATE estimates. In the observational sample, the group with the most negative treatment effects has an average CATE of -0.561 (sd = 0.027), while the group with the smallest absolute effects has an average CATE of -0.414 (sd = 0.025). Appendix Figure \ref{fig:shap_summary_binary} shows SHAP values showing the feature importance in predicting the estimated CATE in each sample. While not identical, the same device characteristics influence the CATE in similar ways across the experimental and observational samples. In other words, the observational sample is not only able to recreate a similar ordering of average CATE across the groups identified in the experimental sample, but also to identify these key splits in the absence of the experimental benchmark.\footnote{Appendix Figure \ref{fig:distilled_cate_trees_binary_obs_only} shows the shallow tree re-estimated using only the observational sample. As expected from the SHAP values, the splits include most of the same features as the experimental tree. } 

Overall, we view this as encouraging evidence that CATE estimators can accurately rank units by treatment effect sizes. This ordering is important for targeting rationed or expensive treatments. Unfortunately, our data sample is not a useful test case for the ability of CATE models to recover the magnitude of heterogeneous treatment effects.

\section{Conclusion} 
\label{sec:conclusion}
This paper provides a new empirical validation study of observational causal inference methods using a unique dataset. Unlike earlier comparisons such as \citet{LaLonde1986}, our clean paired design contains both treated and untreated units in both experimental and observational samples. This richer setup allows for a more direct and comprehensive assessment of observational methods relative to an experimental benchmark.

We find that, when modern best practices are employed—namely careful overlap trimming, AutoML-driven hyperparameter tuning, and model ensembling—observational estimators can closely recover experimental treatment effects. Across a variety of estimators, including outcome modeling, inverse probability weighting, matching, and doubly robust methods, our results show strong agreement with the experimental ground truth for the general performance outcome. These findings extend and reinforce earlier conclusions by \citet{DehejiaWahba1999}, \citet{crump2009dealing}, \citet{xuimbens2025}, \citet{bach2024}, \citet{breiman1996bagging} among others, regarding the critical role of design-stage choices in observational studies. Moreover, when we reanalyze the LaLonde dataset using our proposed workflow, we find similar improvements: applying modern trimming, tuning, and ensembling strategies helps improve alignment between observational and experimental estimates, highlighting the broad applicability of these lessons.


At the same time, our results underscore the fundamental limits of observational causal inference. Even when we successfully recover the experimental benchmark for the general performance outcome, and even when sensitivity analysis provides some credibility to these estimates, this alignment does not in any way \textit{validate} the unconfoundedness assumption, which remains inherently untestable. The analysis of the targeted outcome highlights this challenge starkly: despite applying the same best practices and observing high robustness values, flexible estimators still fail to recover the experimental benchmark. Overall, our findings contribute to the growing literature demonstrating that careful design and validation practices can greatly enhance the credibility of observational causal inference. Yet they also reaffirm that, at its core, causal identification depends as much on substantive assumptions and domain knowledge as on statistical technique. 

\bibliographystyle{apalike}
\bibliography{library}  

\clearpage
\section*{Tables and Figures}
\FloatBarrier

\FloatBarrier

\begin{table}[htbp]
\centering
\caption{Data Sample Features}
\label{tab:data_features}
\begin{tabular}{p{0.22\linewidth} p{0.14\linewidth} p{0.6\linewidth}}
\toprule
Variable & Type & Description \\
\midrule
Treatment & binary & $1:=$ new feature enabled \\
Outcome 1 & binary & metric of device performance \\
Outcome 2 & continuous & metric of device performance \\
\midrule
Total\_Device\_Usage & continuous & Total seconds device was in use over the 2-week period \\
Browser\_x\_Usage & continuous & Seconds spent in an internet browser over the 2-week period, by anonymized browser (i.e. Chrome, Firefox) \\
EngagementSegment & ordinal & Indicator of device's past usage patterns (from heavy use to inactive) \\
A14Region & categorical & Scrambled indicator of geographic region\footnote{In the publicly released data set the geographic labels are real, but have been shuffled. For example, devices labeled "France" may be in Canada and so on.} \\
AppCategoryCohort & categorical & Profile of the device's primary user from past usage patterns across apps (i.e. gamer, worker) \\
Manufacturer & categorical & Anonymized indicator of device manufacturer \\
Device\_Spec\_A & binary & \multirow{3}{*}{Other device characteristics, such as laptop vs desktop} \\
Device\_Spec\_B & categorical &  \\
Device\_Spec\_C & categorical &  \\
\bottomrule
\end{tabular}
\end{table}

\begin{table}[htbp]
\centering
\caption{Optimal Sample Trimming}
\label{tab:trimming_summary}
\begin{tabular}{lccc}
\toprule
 & Untrimmed & Trimmed & \% dropped \\
\midrule
Number of Observations & 445,286 & 304,651 & 31.6\% \\
Number of Treated Units & 20,189 & 18,585 & 7.9\% \\
Number of Control Units & 425,097 & 286,066 & 32.7\% \\
\bottomrule
\end{tabular}
\vspace{0.2cm}
\caption*{\small \textit{Notes:} This table summarizes the results of implementing the \citet{crump2009dealing} trimming procedure on the observational software feature release sample.}
\end{table}

\begin{table}[htbp]\centering
\caption{Summary of First-Stage Model Performance on Whole and Trimmed Samples}
\label{tab:first_stage_summary_combined}
\begin{tabular}{lccc|ccc}
\toprule
 & \multicolumn{3}{c|}{Whole Sample} & \multicolumn{3}{c}{Trimmed Sample} \\
\cmidrule(l{2pt}r{2pt}){2-4} \cmidrule(l{2pt}r{2pt}){5-7}
 & $\hat{\mu}_0$ R$^2$ & $\hat{\mu}_1$ R$^2$ & $\hat{e}$ AUC & $\hat{\mu}_0$ R$^2$ & $\hat{\mu}_1$ R$^2$ & $\hat{e}$ AUC \\
\midrule
\hspace{3pt}Tuned LGBM & 0.150 & 0.127 & 0.688 & 0.130 & 0.126 & 0.597 \\
\hspace{3pt}Tuned RF & 0.138 & 0.124 & 0.687 & 0.123 & 0.121 & 0.593 \\
\hspace{3pt}Default LGBM & 0.150 & 0.114 & 0.686 & 0.130 & 0.115 & 0.592 \\
\hspace{3pt}Logistic/Linear & 0.112 & 0.109 & 0.675 & 0.095 & 0.107 & 0.578 \\
\hspace{3pt}Tuned NN & 0.105 & 0.117 & 0.660 & 0.078 & 0.076 & 0.567 \\
\hspace{3pt}Tuned Lasso & 0.112 & 0.109 & 0.657 & 0.095 & 0.107 & 0.563 \\
\hspace{3pt}Default Lasso & 0.067 & 0.064 & 0.657 & 0.050 & 0.061 & 0.563 \\
\hspace{3pt}Default RF & 0.086 & 0.069 & 0.626 & 0.064 & 0.061 & 0.537 \\
\hspace{3pt}Weakly-tuned NN & 0.022 & 0.029 & 0.510 & 0.031 & 0.039 & 0.498 \\
\bottomrule
\end{tabular}
\vspace{0.2cm}
\caption*{\small \textit{Notes:} This table reports performance metrics for first-stage nuisance models used in the causal analysis. The first two columns report the $R^2$ scores for the outcome models $\hat{\mu}_0(X_i)$ and $\hat{\mu}_1(X_i)$, evaluated on $\{Y_i : D_i=0\}$ and $\{Y_i : D_i=1\}$ respectively. That is, we evaluate our two outcome models' performance based on the subset of observable outcomes on which they were cross-fit and trained. The third column reports the Area Under the Curve (AUC) scores for the propensity score model $\hat{e}(X_i)$. Results are shown separately for the whole sample and the sample after propensity score trimming. Higher scores are indicative of better first-stage nuisance estimation.}
\end{table}

\FloatBarrier

\clearpage
\newpage 
\appendix
\part*{Appendix}

\numberwithin{figure}{section} 
\numberwithin{table}{section} 

\section{Sample Details} 
We assess covariate balance using standardized mean differences (SMD)
\citep{imbens_matching_2015}, calculated as the difference in group
means divided by the square root of the average of the two
within-group sample variances. We show the SMDs between covariates in the observational and experimental samples in Figure \ref{tab:between_sample_smd}. We also shown within-sample SMDs between treated and untreated groups in Figure \ref{tab:within_sample_smd}.  For variable $j$, let $\mathcal{I}_g$ contain the $n_g$ devices in group $g$, let $X_{ij}$ denote device $i$'s value, and let $\bar X_{gj}=n_g^{-1}\sum_{i\in\mathcal{I}_g}X_{ij}$. Define
$$  s_{gj}^{2}
  =\frac{1}{n_g-1}\sum_{i\in\mathcal{I}_g}
    (X_{ij}-\bar X_{gj})^{2}, \qquad \mathrm{SMD}_{j} =\frac{\bar X_{1j}-\bar X_{0j}}{\sqrt{(s_{1j}^{2}+s_{0j}^{2})/2}}.
$$

For Table \ref{tab:between_sample_smd}, group 1 is the observational
sample and group 0 is the experimental sample. For Table \ref{tab:within_sample_smd}, group 1 is treated and group 0 is untreated, with the calculation performed separately in each sample. These are unit-free descriptive comparisons, with larger absolute values indicating greater sample imbalance. Following a common rule of thumb for assessing covariate balance, we bold entries whose reported absolute SMD is at least $0.10$. 

\begin{table}[htbp]
\centering
\caption{Standardized Mean Differences Between Observed and Experimental Samples}
\label{tab:between_sample_smd}
\footnotesize
\begin{adjustbox}{max totalsize={\textwidth}{\textheight},center}
\begin{tabular}{lccc}
\toprule
Variable & Observed Mean & Experiment Mean & SMD \\
\midrule
Treatment & 0.045 (0.208) & 0.026 (0.160) & \textbf{0.102} \\
Targeted Outcome & 0.962 (0.192) & 0.976 (0.154) & -0.080 \\
General Performance Outcome & 4.116 (6.878) & 4.017 (6.666) & 0.015 \\
\midrule
Total\_Device\_Usage & 138,974 (139,359) & 136,247 (138,453) & 0.020 \\
Browser\_1\_Usage & 22,238 (57,848) & 21,805 (57,119) & 0.008 \\
Browser\_2\_Usage & 41,787 (79,143) & 40,944 (78,301) & 0.011 \\
Browser\_3\_Usage & 4,661 (28,999) & 4,639 (29,169) & 0.001 \\
Browser\_4\_Usage & 38.388 (1,461) & 31.971 (1,159) & 0.005 \\
Other\_Browser\_Usage & 3,693 (26,972) & 3,549 (26,444) & 0.005 \\
EngagementSegment\_General & 0.112 (0.316) & 0.112 (0.316) & -0.001 \\
EngagementSegment\_Highly Engaged & 0.460 (0.498) & 0.450 (0.497) & 0.021 \\
EngagementSegment\_Inactive & 0.003 (0.057) & 0.003 (0.058) & -0.001 \\
EngagementSegment\_Low Engaged & 0.145 (0.352) & 0.151 (0.358) & -0.017 \\
A14Region\_Korea & 0.041 (0.199) & 0.041 (0.199) & -0.000 \\
A14Region\_United States & 0.026 (0.158) & 0.026 (0.160) & -0.003 \\
A14Region\_Japan & 0.017 (0.129) & 0.017 (0.129) & -0.000 \\
A14Region\_India & 0.030 (0.171) & 0.030 (0.169) & 0.003 \\
A14Region\_Western Europe & 0.029 (0.169) & 0.030 (0.171) & -0.004 \\
A14Region\_UK & 0.124 (0.329) & 0.123 (0.329) & 0.001 \\
A14Region\_Greater China & 0.048 (0.214) & 0.047 (0.212) & 0.004 \\
A14Region\_Canada & 0.175 (0.380) & 0.175 (0.380) & -0.001 \\
A14Region\_Latam & 0.015 (0.123) & 0.015 (0.123) & -0.000 \\
A14Region\_France & 0.073 (0.260) & 0.072 (0.259) & 0.003 \\
A14Region\_Germany & 0.033 (0.178) & 0.032 (0.176) & 0.005 \\
A14Region\_APAC & 0.024 (0.152) & 0.024 (0.153) & -0.002 \\
A14Region\_CEE & 0.274 (0.446) & 0.275 (0.446) & -0.002 \\
A14Region\_MEA & 0.068 (0.251) & 0.069 (0.253) & -0.003 \\
AppCategoryCohort\_Developer & 0.036 (0.187) & 0.035 (0.184) & 0.007 \\
AppCategoryCohort\_Gamer & 0.105 (0.307) & 0.104 (0.305) & 0.004 \\
AppCategoryCohort\_General & 0.112 (0.316) & 0.112 (0.316) & -0.001 \\
AppCategoryCohort\_Media & 0.026 (0.159) & 0.026 (0.160) & -0.002 \\
AppCategoryCohort\_Productivity & 0.170 (0.376) & 0.172 (0.377) & -0.003 \\
Manufacturer\_1 & 0.116 (0.320) & 0.115 (0.319) & 0.003 \\
Manufacturer\_2 & 0.137 (0.343) & 0.136 (0.342) & 0.003 \\
Manufacturer\_3 & 0.212 (0.409) & 0.210 (0.408) & 0.004 \\
Manufacturer\_4 & 0.166 (0.372) & 0.166 (0.372) & -0.001 \\
Manufacturer\_5 & 0.074 (0.262) & 0.076 (0.264) & -0.005 \\
Manufacturer\_6 & 0.219 (0.413) & 0.221 (0.415) & -0.005 \\
Device\_Spec\_A & 0.244 (0.429) & 0.246 (0.431) & -0.006 \\
Device\_Spec\_B\_1 & 0.627 (0.484) & 0.625 (0.484) & 0.004 \\
Device\_Spec\_B\_2 & 0.186 (0.389) & 0.187 (0.390) & -0.005 \\
Device\_Spec\_C\_1 & 0.040 (0.197) & 0.042 (0.200) & -0.007 \\
Device\_Spec\_C\_2 & 0.518 (0.500) & 0.511 (0.500) & 0.015 \\
Device\_Spec\_C\_3 & 0.007 (0.084) & 0.007 (0.085) & -0.001 \\
\bottomrule
\end{tabular}
\end{adjustbox}
\par\smallskip
\begin{minipage}{\textwidth}
\footnotesize\raggedright
\textit{Notes:} This table shows covariate balance between the observational and experimental samples. Standard deviations are  in parentheses. Standardized mean
differences (SMD) are shown on the right-hand column 
\citep{imbens_matching_2015}. Boldface flags reported
$|\mathrm{SMD}|\geq 0.10$, a common diagnostic rule of thumb.
\end{minipage}
\end{table}

\begin{table}[htbp]
\centering
\caption{Within-Sample Standardized Mean Differences Between Treated and Untreated Groups}
\label{tab:within_sample_smd}
\resizebox{\textwidth}{!}{%
\begin{tabular}{lcccccc}
\toprule
& \multicolumn{3}{c}{Observed Sample ($n=445286$)} & \multicolumn{3}{c}{Experiment Sample ($n=435170$) } \\ 
Variable & Treated Mean & Untreated Mean & SMD & Treated Mean & Untreated Mean & SMD \\
\midrule
Targeted Outcome & 0.454 (0.498) & 0.986 (0.119) & \textbf{-1.470} & 0.601 (0.490) & 0.986 (0.118) & \textbf{-1.080} \\
General Performance Outcome & 5.269 (7.983) & 4.061 (6.816) & \textbf{0.163} & 4.438 (6.912) & 4.005 (6.659) & 0.064 \\
\midrule
Total\_Device\_Usage & 194,293 (142,868) & 136,347 (138,642) & \textbf{0.412} & 146,718 (139,427) & 135,964 (138,416) & 0.077 \\
Browser\_1\_Usage & 31,786 (70,701) & 21,784 (57,126) & \textbf{0.156} & 23,077 (59,126) & 21,770 (57,064) & 0.022 \\
Browser\_2\_Usage & 59,955 (94,314) & 40,924 (78,245) & \textbf{0.220} & 44,301 (80,972) & 40,853 (78,225) & 0.043 \\
Browser\_3\_Usage & 6,426 (34,542) & 4,577 (28,707) & 0.058 & 5,528 (33,460) & 4,615 (29,043) & 0.029 \\
Browser\_4\_Usage & 60.601 (1,640) & 37.333 (1,452) & 0.015 & 39.280 (1,144) & 31.773 (1,159) & 0.007 \\
Other\_Browser\_Usage & 5,996 (34,885) & 3,583 (26,532) & 0.078 & 3,272 (22,746) & 3,557 (26,537) & -0.012 \\
EngagementSegment\_General & 0.106 (0.308) & 0.112 (0.316) & -0.021 & 0.110 (0.313) & 0.112 (0.316) & -0.008 \\
EngagementSegment\_Highly Engaged & 0.694 (0.461) & 0.449 (0.497) & \textbf{0.511} & 0.495 (0.500) & 0.449 (0.497) & 0.093 \\
EngagementSegment\_Inactive & 0.001 (0.025) & 0.003 (0.059) & -0.062 & 0.002 (0.042) & 0.003 (0.059) & -0.033 \\
EngagementSegment\_Low Engaged & 0.019 (0.136) & 0.151 (0.358) & \textbf{-0.487} & 0.115 (0.319) & 0.152 (0.359) & \textbf{-0.110} \\
A14Region\_Korea & 0.044 (0.204) & 0.041 (0.199) & 0.012 & 0.046 (0.209) & 0.041 (0.199) & 0.022 \\
A14Region\_United States & 0.028 (0.164) & 0.026 (0.158) & 0.013 & 0.026 (0.158) & 0.026 (0.160) & -0.004 \\
A14Region\_Japan & 0.017 (0.130) & 0.017 (0.129) & 0.003 & 0.016 (0.125) & 0.017 (0.129) & -0.009 \\
A14Region\_India & 0.032 (0.177) & 0.030 (0.170) & 0.015 & 0.031 (0.175) & 0.029 (0.169) & 0.012 \\
A14Region\_Western Europe & 0.026 (0.160) & 0.030 (0.169) & -0.020 & 0.029 (0.169) & 0.030 (0.171) & -0.004 \\
A14Region\_UK & 0.143 (0.351) & 0.123 (0.328) & 0.061 & 0.143 (0.350) & 0.123 (0.328) & 0.058 \\
A14Region\_Greater China & 0.051 (0.221) & 0.048 (0.213) & 0.017 & 0.054 (0.225) & 0.047 (0.211) & 0.031 \\
A14Region\_Canada & 0.166 (0.372) & 0.175 (0.380) & -0.024 & 0.166 (0.372) & 0.175 (0.380) & -0.023 \\
A14Region\_Latam & 0.012 (0.111) & 0.015 (0.123) & -0.026 & 0.014 (0.118) & 0.015 (0.123) & -0.011 \\
A14Region\_France & 0.083 (0.276) & 0.072 (0.259) & 0.039 & 0.073 (0.260) & 0.072 (0.259) & 0.004 \\
A14Region\_Germany & 0.036 (0.185) & 0.033 (0.178) & 0.015 & 0.031 (0.173) & 0.032 (0.176) & -0.006 \\
A14Region\_APAC & 0.023 (0.149) & 0.024 (0.152) & -0.007 & 0.024 (0.155) & 0.024 (0.153) & 0.003 \\
A14Region\_CEE & 0.247 (0.431) & 0.275 (0.447) & -0.063 & 0.257 (0.437) & 0.275 (0.447) & -0.042 \\
A14Region\_MEA & 0.068 (0.252) & 0.068 (0.251) & 0.002 & 0.069 (0.253) & 0.069 (0.253) & 0.000 \\
AppCategoryCohort\_Developer & 0.049 (0.216) & 0.036 (0.186) & 0.066 & 0.041 (0.199) & 0.035 (0.183) & 0.035 \\
AppCategoryCohort\_Gamer & 0.129 (0.335) & 0.104 (0.306) & 0.077 & 0.113 (0.317) & 0.104 (0.305) & 0.030 \\
AppCategoryCohort\_General & 0.106 (0.308) & 0.112 (0.316) & -0.021 & 0.110 (0.313) & 0.112 (0.316) & -0.008 \\
AppCategoryCohort\_Media & 0.019 (0.136) & 0.026 (0.160) & -0.050 & 0.026 (0.159) & 0.026 (0.160) & -0.003 \\
AppCategoryCohort\_Productivity & 0.137 (0.344) & 0.172 (0.377) & -0.097 & 0.160 (0.366) & 0.172 (0.377) & -0.033 \\
Manufacturer\_1 & 0.123 (0.329) & 0.115 (0.320) & 0.024 & 0.116 (0.321) & 0.115 (0.319) & 0.005 \\
Manufacturer\_2 & 0.141 (0.348) & 0.136 (0.343) & 0.013 & 0.138 (0.345) & 0.136 (0.342) & 0.007 \\
Manufacturer\_3 & 0.218 (0.413) & 0.212 (0.409) & 0.016 & 0.203 (0.402) & 0.211 (0.408) & -0.018 \\
Manufacturer\_4 & 0.173 (0.378) & 0.166 (0.372) & 0.020 & 0.162 (0.368) & 0.166 (0.373) & -0.013 \\
Manufacturer\_5 & 0.045 (0.207) & 0.076 (0.264) & \textbf{-0.129} & 0.069 (0.254) & 0.076 (0.265) & -0.025 \\
Manufacturer\_6 & 0.214 (0.410) & 0.219 (0.414) & -0.012 & 0.230 (0.421) & 0.221 (0.415) & 0.022 \\
Device\_Spec\_A & 0.183 (0.387) & 0.247 (0.431) & \textbf{-0.156} & 0.233 (0.422) & 0.247 (0.431) & -0.033 \\
Device\_Spec\_B\_1 & 0.656 (0.475) & 0.625 (0.484) & 0.064 & 0.641 (0.480) & 0.625 (0.484) & 0.035 \\
Device\_Spec\_B\_2 & 0.135 (0.342) & 0.188 (0.391) & \textbf{-0.144} & 0.179 (0.383) & 0.188 (0.390) & -0.023 \\
Device\_Spec\_C\_1 & 0.028 (0.164) & 0.041 (0.198) & -0.073 & 0.043 (0.203) & 0.042 (0.200) & 0.007 \\
Device\_Spec\_C\_2 & 0.671 (0.470) & 0.511 (0.500) & \textbf{0.331} & 0.537 (0.499) & 0.510 (0.500) & 0.055 \\
Device\_Spec\_C\_3 & 0.004 (0.066) & 0.007 (0.085) & -0.037 & 0.008 (0.089) & 0.007 (0.084) & 0.009 \\
\bottomrule
\end{tabular}}
\par\smallskip
\begin{minipage}{\textwidth}
\footnotesize\raggedright
\textit{Notes:} This table shows covariate balance within the treated and untreated groups in the observational and experimental samples. Standard deviations are  in parentheses. Standardized mean
differences (SMD) are shown on the right-hand column 
\citep{imbens_matching_2015}. Boldface flags reported
$|\mathrm{SMD}|\geq 0.10$, a common diagnostic rule of thumb.
\end{minipage}
\end{table}

\paragraph{Control-sample placebo check.}
As an additional check on comparability, we compare untreated devices from
the experimental and observational samples under the same sample eligibility
criteria. Experimental controls form the pseudo-treated group and
observational controls the comparison group; neither received the feature.
We estimate the experimental-minus-observational contrast using cross-fitted
augmented inverse-probability weighting. Table~\ref{tab:control-placebo} shows no statistically detectable
difference for either outcome. These findings are consistent with
comparability of the untreated samples in the analysis population.

\begin{table}[htbp]
\centering
\caption{Placebo tests between experimental and observational controls}
\label{tab:control-placebo}
\small
\begin{tabular}{lrrrr}
\hline
Outcome & Estimate & SE & 95\% CI & $p$-value \\
\hline
General performance & $0.0050$ & $0.0133$
  & $[-0.0211,\;0.0311]$ & $0.707$ \\
Targeted & $0.0035$ & $0.0249$
  & $[-0.0454,\;0.0524]$ & $0.890$ \\
\hline
\end{tabular}
\par\smallskip
\begin{minipage}{\textwidth}
\footnotesize\raggedright
\textit{Notes:} The placebo tests compare outcomes for 423,695
experimental controls and 424,483 observational controls, adjusting
for observed covariates. Because the feature is disabled in both
groups, these comparisons provide an outcome-based check of sample
comparability, complementing the covariate balance checks.
Neither difference is statistically distinguishable from zero.
\end{minipage}
\end{table}
\FloatBarrier

\section{Propensity Estimation and Tuning} 
Here we provide the propensity score distributions across a range of different flexible and inflexible models. We compare the ``untuned'' or default sklearn implementations, as well as an AutoML-tuned version.\footnote{Details on default parameters can be found in the API pages for
\href{https://scikit-learn.org/stable/api/sklearn.linear\_model.html}{linear models},
\href{https://scikit-learn.org/stable/api/sklearn.ensemble.html}{random forests}, and
\href{https://lightgbm.readthedocs.io/en/latest/Python-API.html\#scikit-learn-api}{boosted trees}.}
 For the neural network, we distinguish between a \emph{fully-tuned} neural network and a \emph{weakly-tuned} baseline. Both are optimized with \texttt{Optuna} (another Bayesian optimization AutoML framework), but the search space for the fully-tuned network is much deeper and more expressive. Details are given in Appendix Table \ref{tab:nn_tuning_spaces}. 
 
\begin{table}[htbp]
\centering
\caption{Hyper-parameter search spaces for the fully-tuned and the weakly-tuned neural networks.}
\label{tab:nn_tuning_spaces}
\begin{tabular}{@{}p{4cm}p{7cm}p{5.5cm}@{}}
\toprule
\textbf{Model} & \textbf{Fully-tuned} & \textbf{Weakly-tuned} \\
\midrule
Hidden layers $n_{\text{layers}}$    & $\{2,\dots,7\}$                   & $\{1,2,3\}$ \\[2pt]
Units per layer $n^{(l)}_{\text{units}}$ & $\{32,64,\dots,1024\}$  & $\{16,32,\dots,128\}$ \\[2pt]
Activation functions                 & \texttt{relu}, \texttt{tanh}, \texttt{gelu}, \texttt{selu}, \texttt{leaky\_relu}, \texttt{elu} & \texttt{relu}, \texttt{tanh}, \texttt{gelu} \\[2pt]
Dropout rate                         & $[0,0.5]$            & $[0,0.3]$  \\[2pt]
Batch normalization                  & \texttt{True} / \texttt{False}    &  \texttt{False} (Fixed) \\[2pt]
Skip connections                     & \texttt{True} / \texttt{False}    & \texttt{False} (Fixed) \\[2pt]
Learning rate                      & $[10^{-4},\,5\times10^{-2}]$  & $[10^{-4},\,5\times10^{-2}]$ \\[2pt]
Weight decay ($\ell_2$)              & $[0,\,5\times10^{-2}]$            & $0$ (Fixed) \\[2pt]
\bottomrule
\end{tabular}
\end{table}

\begin{figure}[htbp]
    \centering
    \includegraphics[width=\linewidth]{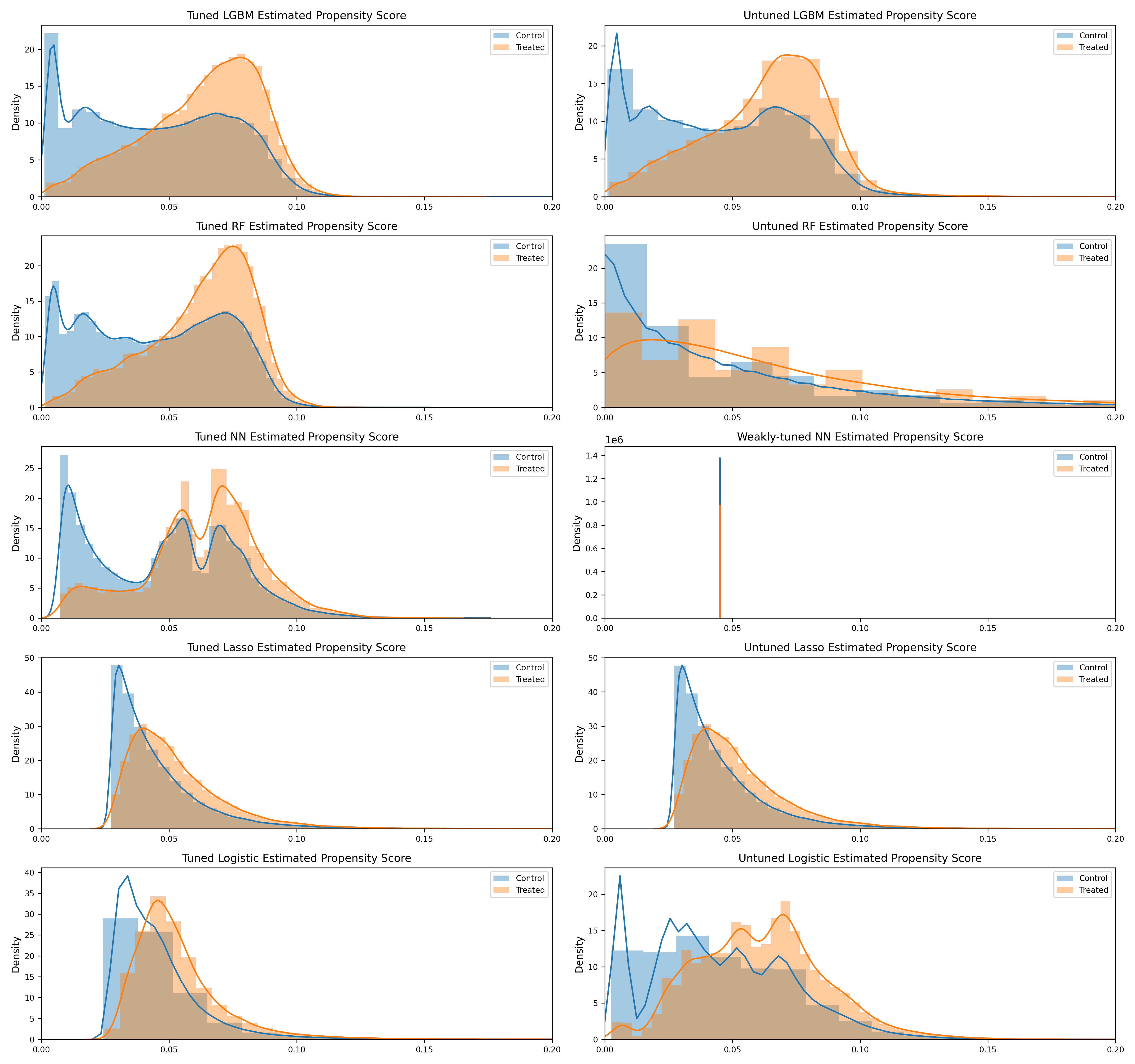}
    \caption{Untrimmed Sample Propensity Scores: Model Comparison with and without Tuning}
    \label{fig:Model_untrimmed_Comparison_PScore_Dist}
\end{figure}

\begin{figure}[htbp]
    \centering
    \includegraphics[width=\linewidth]{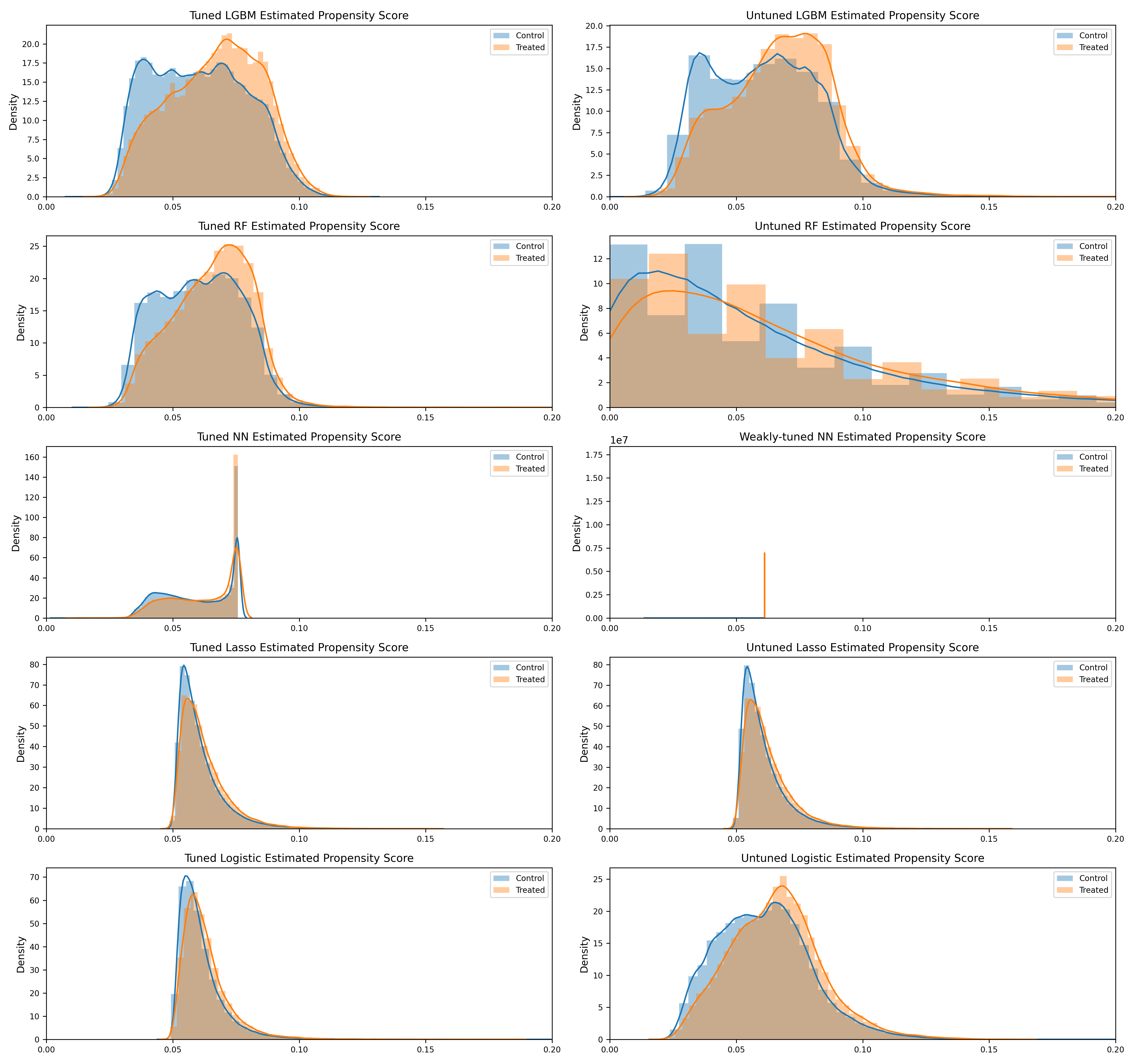}
    \caption{Trimmed Sample Propensity Scores: Model Comparison with and without Tuning}
    \label{fig:Model_trimmed_Comparison_PScore_Dist}
\end{figure}

\begin{figure}[htbp]
    \centering
    \begin{minipage}{0.48\textwidth}
        \centering
        \includegraphics[width=\linewidth]{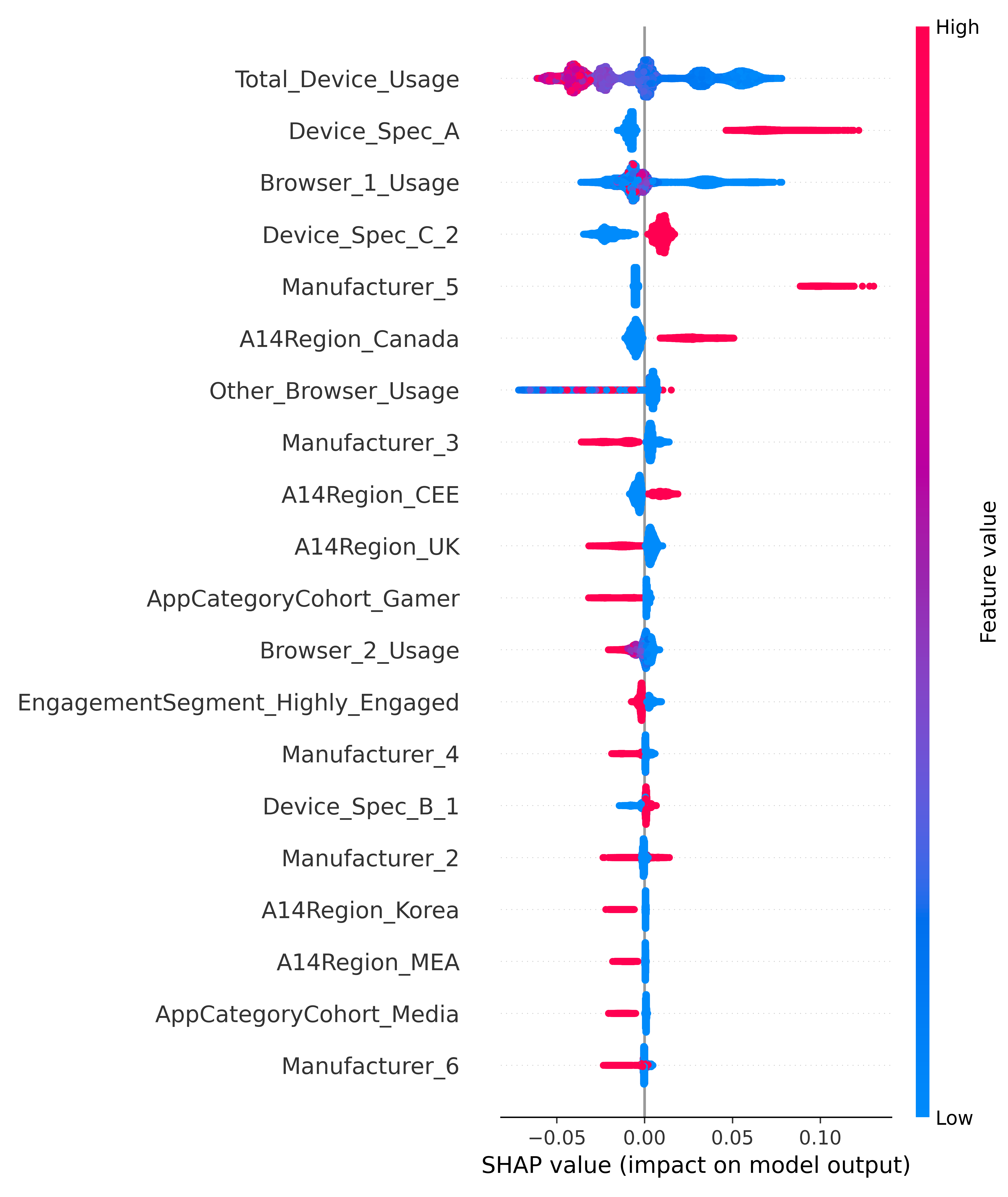}
        \caption*{(a) Experimental Sample: SHAP Summary}
    \end{minipage}\hfill
    \begin{minipage}{0.48\textwidth}
        \centering
        \includegraphics[width=\linewidth]{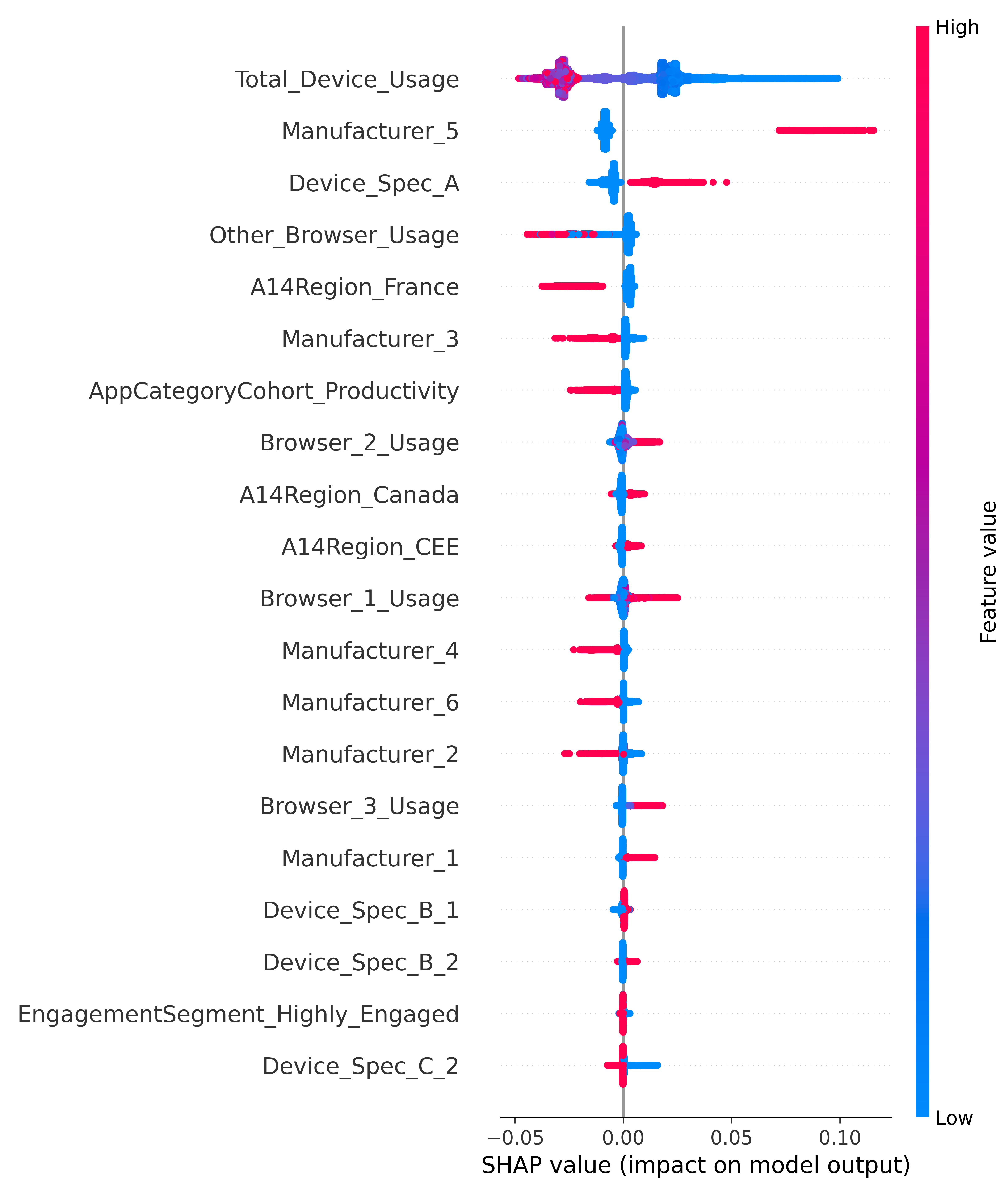}
        \caption*{(b) Observational Sample: SHAP Summary}
    \end{minipage}
    \caption{SHAP summary plots of feature importance in Q-aggregated CATE models for the targeted outcome. Consistent signals are observed in both datasets.}
    \label{fig:shap_summary_binary}
\end{figure}

\FloatBarrier

\section{Additional ATE Results on the Untrimmed Samples}

\begin{figure}[htbp]
    \centering
    \includegraphics[width=0.85\linewidth]{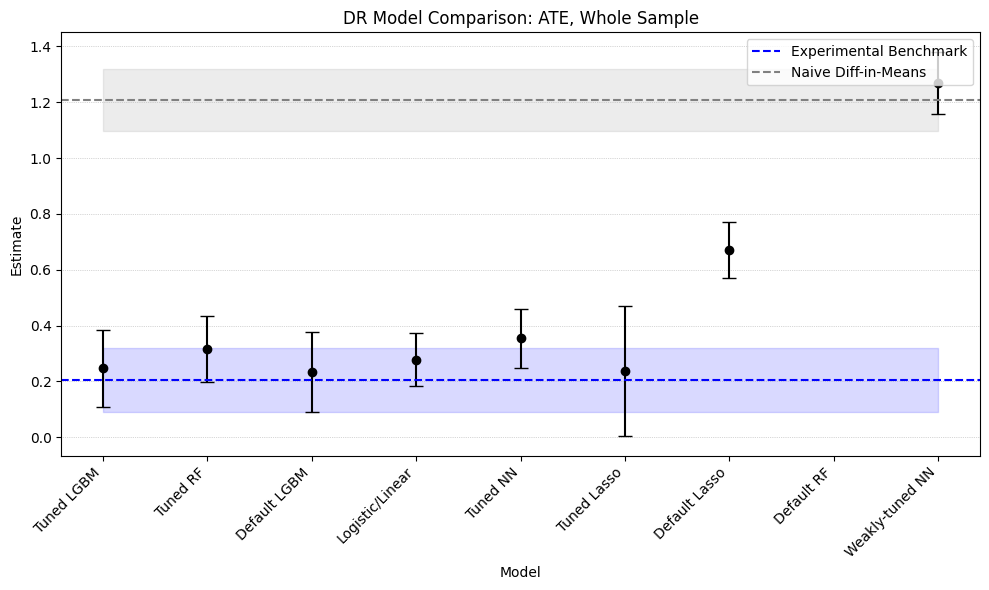}
    \caption{\textbf{DR Estimates Across Different First-Stage Models (Untrimmed Sample):} This figure compares ATE estimates from doubly robust estimators using a range of flexible and rigid first-stage models for outcome and treatment modeling. Both tuned and default sklearn versions are considered. Poorly tuned flexible models, particularly RFs and NNs, exhibit severe bias and instability, highlighting the critical role of hyperparameter tuning in these cases. The default random forest implementation yields an estimate of -93.776 (67.239), driven by very unstable propensity weights, and is thus not pictured in the figure. Trimming is not applied here, amplifying model misspecification and overlap issues.}
    \label{fig:dr_comparison_untrimmed}
\end{figure}

\FloatBarrier

\section{ATT Results} 

\begin{figure}[htbp]
    \centering
    \begin{subfigure}[t]{0.8\textwidth}
        \centering
        \includegraphics[width=\textwidth]{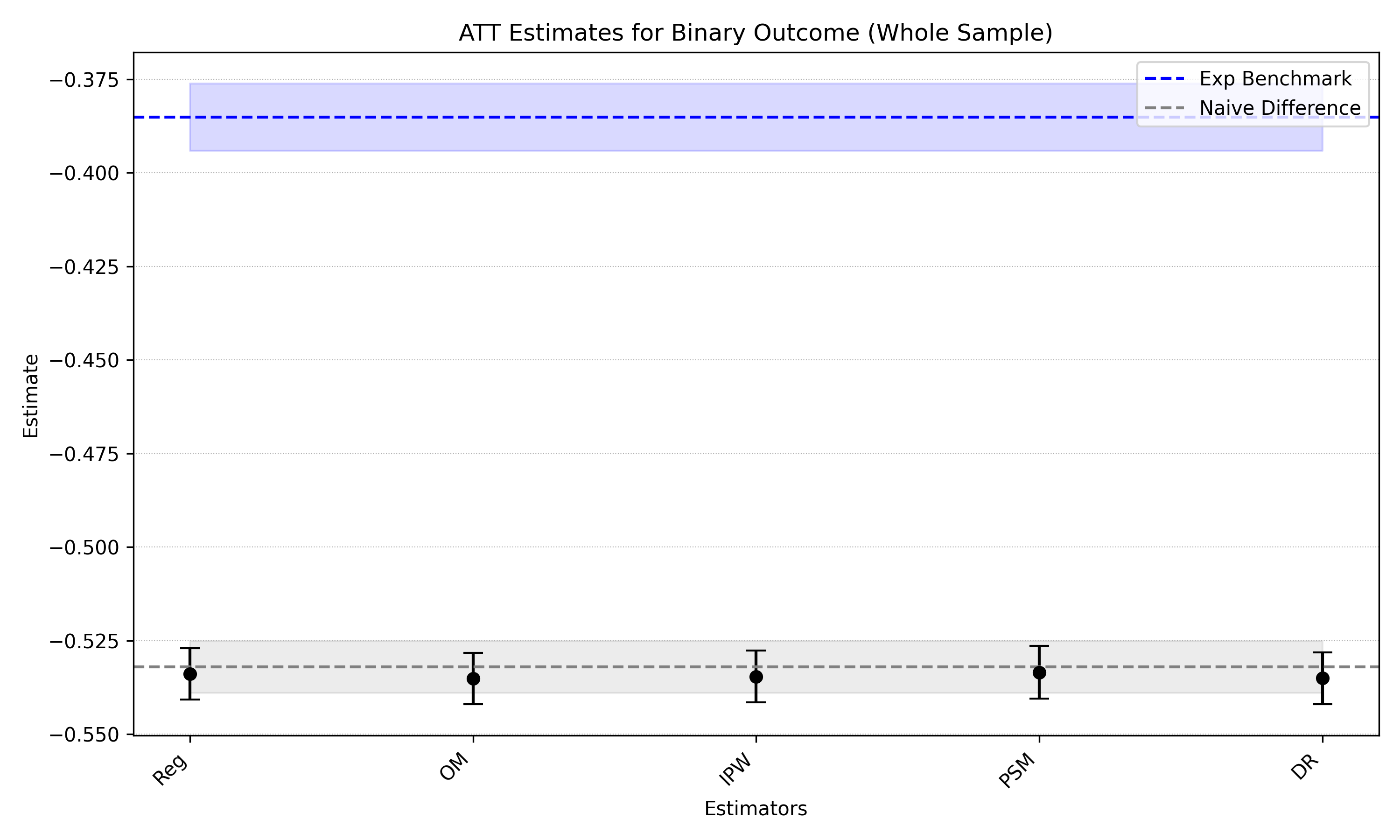}
        \caption{Untrimmed Sample}
        \label{fig:att_summary_binary_untrimmed}
    \end{subfigure}

    \vspace{0.5em}

    \begin{subfigure}[t]{0.8\textwidth}
        \centering
        \includegraphics[width=\textwidth]{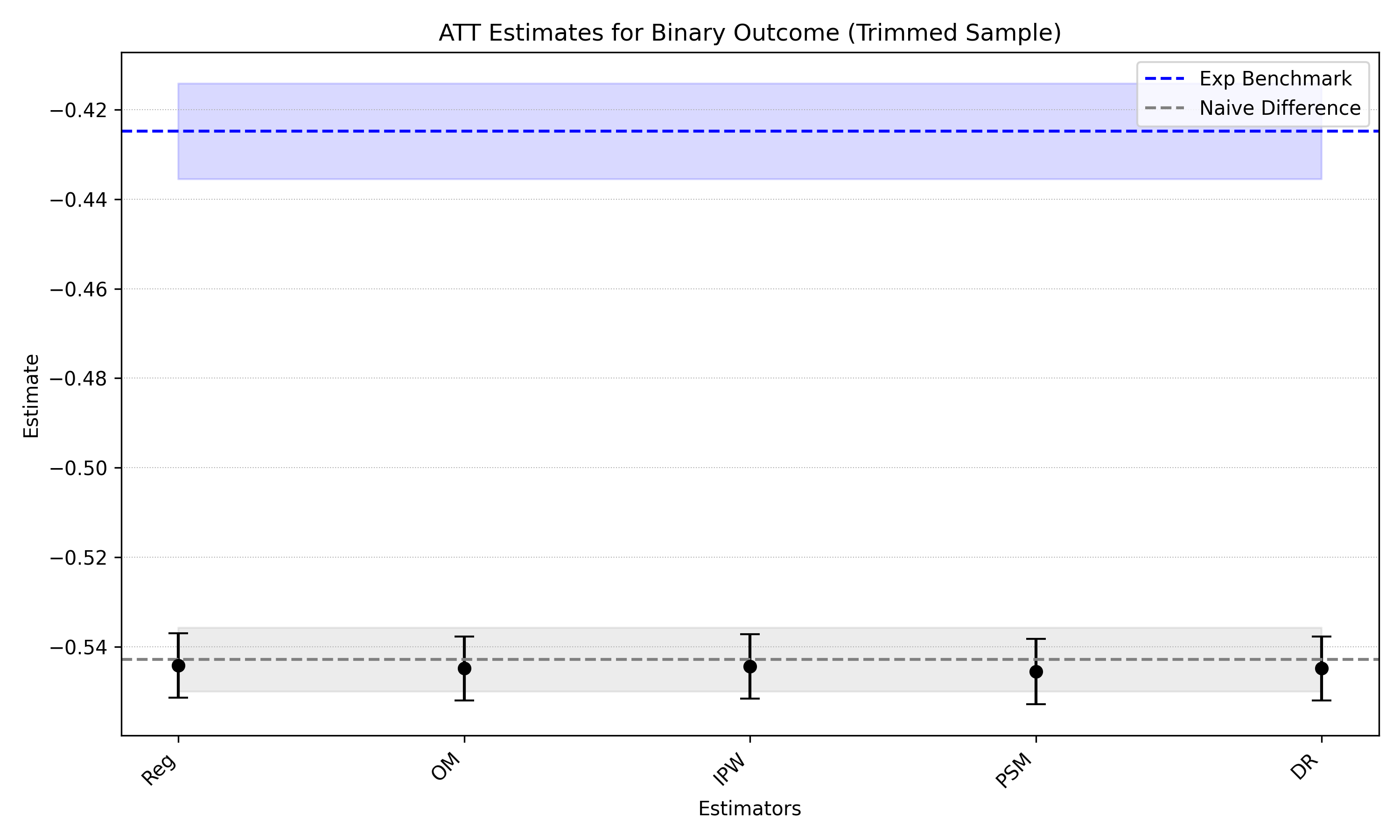}
        \caption{Trimmed Sample}
        \label{fig:att_summary_binary_trimmed}
    \end{subfigure}

    \caption{\textbf{ATT Estimates for Targeted Outcome:} Panels (a) and (b) display ATT estimates and their 95\% confidence intervals for five estimators: regression adjustment (Reg), outcome modeling (OM), inverse probability weighting (IPW), propensity score matching (PSM), and doubly robust estimation (DR). Panel (a) shows results from the untrimmed sample, and panel (b) presents estimates after overlap trimming. Across both samples, observational estimators fail to recover the experimental benchmark (blue dashed line), with estimates consistently biased downward and concentrated around the naive difference in means (gray dashed line). This pattern strongly suggests unobserved confounding. Unlike the general performance outcome, where modern econometric advancements help align with the experimental benchmark, these results show that even with doubly robust methods and trimming, we fail to recover the ground truth.}
    \label{fig:att_summary_binary}
\end{figure}

\begin{figure}[htbp]
    \centering
    \begin{subfigure}[t]{0.8\textwidth}
        \centering
        \includegraphics[width=\textwidth]{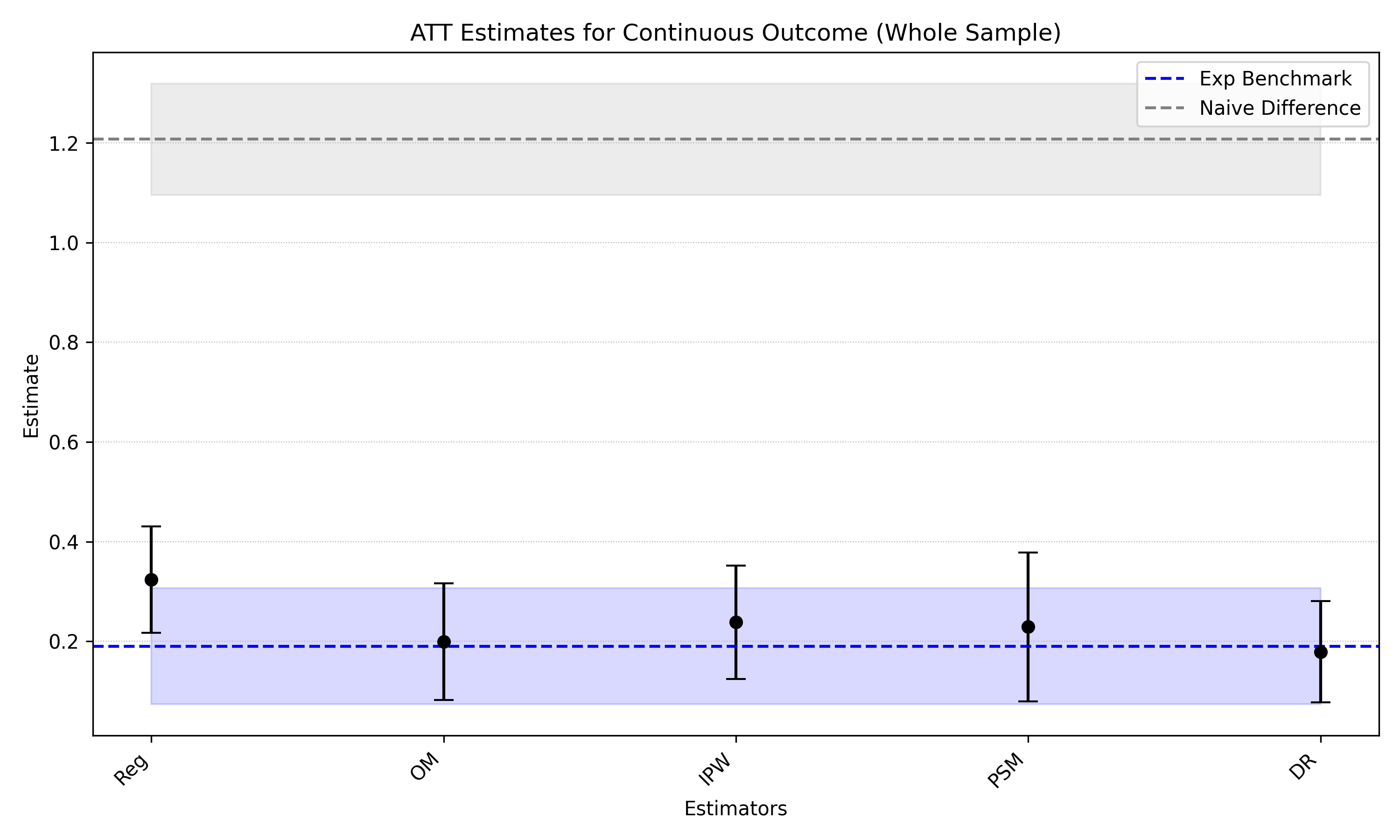}
        \caption{Untrimmed Sample}
        \label{fig:att_summary_untrimmed}
    \end{subfigure}

    \vspace{0.5em}

    \begin{subfigure}[t]{0.8\textwidth}
        \centering
        \includegraphics[width=\textwidth]{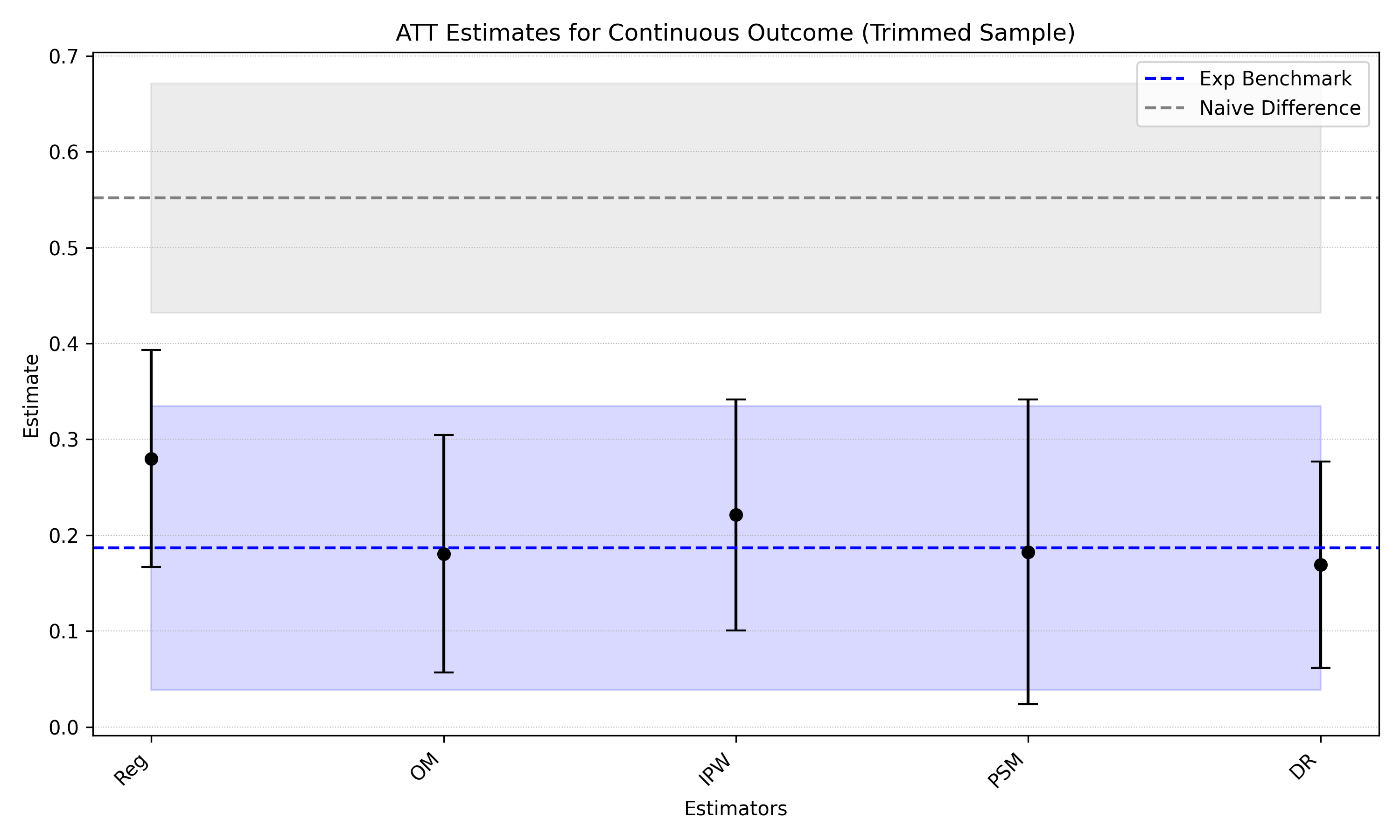}
        \caption{Trimmed Sample}
        \label{fig:att_summary_trimmed}
    \end{subfigure}

    \caption{\textbf{ATT Estimates for General Performance Outcome:} Panels (a) and (b) display ATT estimates and their 95\% confidence intervals for five different estimators: regression adjustment (Reg), outcome modeling (OM), inverse probability weighting (IPW), propensity score matching (PSM), and doubly robust estimation (DR). Panel (a) shows the full untrimmed sample; Panel (b) displays estimates from the trimmed sample. In each plot, the black dashed line and shaded band indicate the naive difference in means and its 95\% confidence interval, while the blue dashed line and shaded region denote the experimental benchmark and its 95\% CI. Trimming improves estimator agreement with the benchmark and reduces variance. These results are consistent with findings in \citet{DehejiaWahba1999}, \citet{crump2009dealing}, and \citet{xuimbens2025}.}
    \label{fig:att_summary_continuous}
\end{figure}

We show average treatment effect on the treated results for the targeted outcome in Figure \ref{fig:att_summary_binary}, and those for the general performance outcome in Figure \ref{fig:att_summary_continuous}.

\FloatBarrier

\section{Additional CATE Results}

\begin{figure}[htbp]
    \centering
    \includegraphics[width=0.8\linewidth]{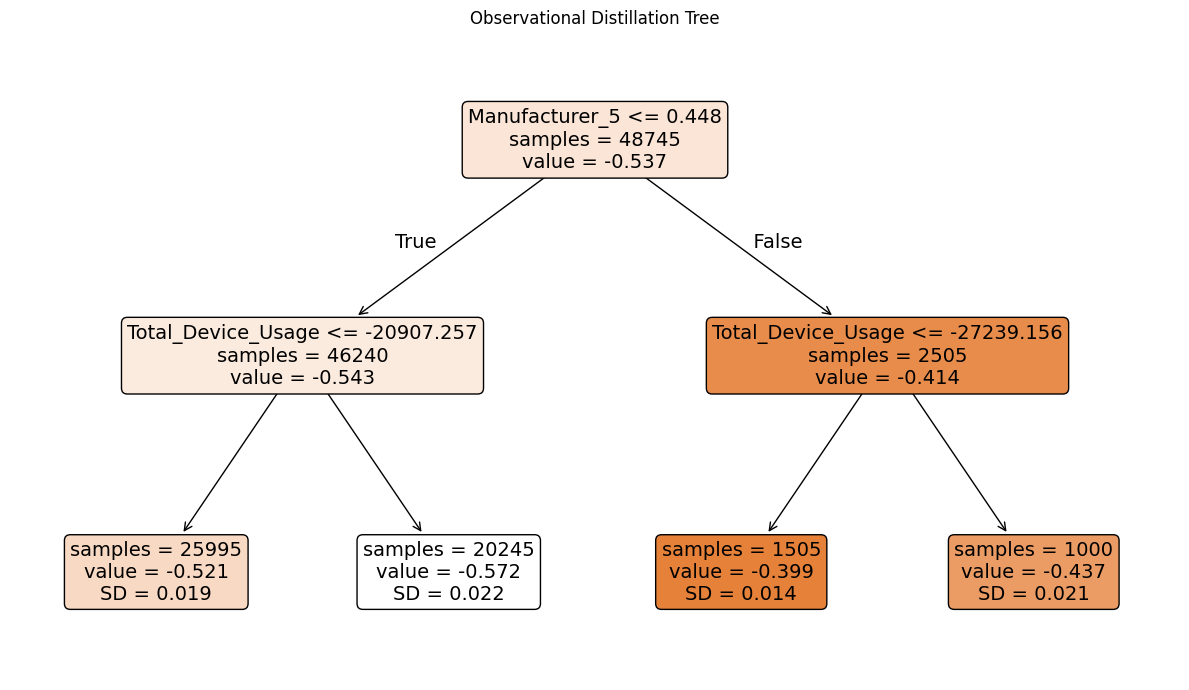}
    \caption{Distilled CATE tree from a Q-aggregated model for the targeted outcome, estimated using only the observational sample. This tree recovers many of the same splits as the experimental tree like Total\_Device\_Usage and Manufacturer\_5. 
    Each final leaf shows the standard deviation of estimated CATE within the group (not the standard error of the estimate).}
    \label{fig:distilled_cate_trees_binary_obs_only}
\end{figure}

\FloatBarrier

\section{Experimental Adjustment}
We apply a doubly-robust adjustment on the experiment to get the benchmark.\footnote{We note a simple regression adjustment leads to an almost indistinguishable benchmark.} The summary of this is given in Appendix Table \ref{tab:dr_summary}. 

\begin{table}[htbp]
\centering
\begin{threeparttable}
\caption{Treatment Effect Estimates Under Trimming and Estimation Methods}
\label{tab:dr_summary}
\begin{tabular}{lcccc}
\toprule
\textbf{Sample} & \textbf{Estimator} & \textbf{Full Sample} & \textbf{Trimmed} & \textbf{Trimmed Off-Support} \\
\midrule
\multirow{3}{*}{Experimental} 
& Difference-in-Means & 0.433 (0.065) & 0.309 (0.082) & 0.236 (0.086) \\
& DR (ATT) & 0.190 (0.059) & 0.187 (0.076) & 0.184 (0.081) \\
& DR (ATE) & 0.205 (0.059) & 0.210 (0.077) & 0.165 (0.083) \\
\midrule
\multirow{3}{*}{Observational} 
& Difference-in-Means & 1.208 (0.057) & 0.552 (0.061) & 0.605 (0.149) \\
& DR (ATT) & 0.179 (0.052) & 0.169 (0.055) & 0.317 (0.144) \\
& DR (ATE) & 0.247 (0.061) & 0.206 (0.056) & 0.414 (0.172) \\
\bottomrule
\end{tabular}
\begin{tablenotes}
\small
\item \textit{Notes:} Each entry reports the estimated treatment effect with its standard error in parentheses. ``Trimmed Off-Support'' denotes the excluded region from \citet{crump2009dealing} trimming.
\end{tablenotes}
\end{threeparttable}
\end{table}

\FloatBarrier

\section{More on Sensitivity Analysis}
\subsection{Benchmarking} 
We report our benchmarking $(C_Y, C_D, \rho)$ for each observed covariate in Appendix Table \ref{tab:covariate-benchmarking}
\begin{table}[htbp]
    \centering
    \begin{subtable}[t]{0.44\textwidth}
        \centering
        \scriptsize
        \caption{General performance outcome}
        \begin{tabular}{lccc}
        \toprule
         & $C_Y$ & $C_D$ & $\rho$ \\
        \midrule
        Total\_Device\_Usage & 0.0232 & 0.013 & 0.1038 \\
        Other\_Browser\_Usage & 0.0111 & 0.001 & 0.0675 \\
        Manufacturer\_2 & 0.0059 & 0.0009 & 0.0519 \\
        Browser\_2\_Usage & 0.0053 & 0 & $-1$ \\
        A14Region\_UK & 0.0033 & 0.0008 & 0.0761 \\
        A14Region\_Latam & 0.0029 & 0.0012 & $-0.1385$ \\
        Browser\_1\_Usage & 0.0026 & 0 & $-1$ \\
        AppCategoryCohort\_Gamer & 0.0015 & 0 & 1 \\
        Manufacturer\_4 & 0.0015 & 0.0011 & 0.0027 \\
        Device\_Spec\_C\_2 & 0.0013 & 0.0033 & 0.5336 \\
        AppCategoryCohort\_Developer & 0.0013 & 0 & 0.176 \\
        Manufacturer\_3 & 0.0012 & 0.0008 & $-0.1088$ \\
        Device\_Spec\_B\_1 & 0.0008 & 0.0003 & $-0.1928$ \\
        A14Region\_Canada & 0.0008 & 0.0011 & $-0.0921$ \\
        Manufacturer\_6 & 0.0007 & 0 & $-1$ \\
        Device\_Spec\_C\_1 & 0.0005 & 0.0008 & $-0.0018$ \\
        Manufacturer\_1 & 0.0004 & 0 & $-1$ \\
        Browser\_3\_Usage & 0.0004 & 0.0001 & 0.0821 \\
        EngagementSegment\_Highly Engaged & 0.0004 & 0.0151 & 0.2069 \\
        Browser\_4\_Usage & 0.0004 & 0.0006 & $-0.0241$ \\
        Device\_Spec\_A & 0.0004 & 0.0013 & 0.2863 \\
        Manufacturer\_5 & 0.0003 & 0 & $-1$ \\
        A14Region\_Greater China & 0.0002 & 0.0006 & $-0.0563$ \\
        Device\_Spec\_C\_3 & 0.0002 & 0.0012 & $-0.081$ \\
        AppCategoryCohort\_Media & 0.0002 & 0 & 1 \\
        Device\_Spec\_B\_2 & 0.0001 & 0.0005 & $-0.4256$ \\
        A14Region\_Korea & 0.0001 & 0.0011 & $-0.3347$ \\
        EngagementSegment\_General & 0.0001 & 0.0014 & $-0.1758$ \\
        A14Region\_Germany & 0.0001 & 0.0011 & $-0.0069$ \\
        A14Region\_India & 0 & 0.001 & $-0.2162$ \\
        A14Region\_MEA & 0 & 0.001 & $-0.197$ \\
        A14Region\_CEE & 0 & 0.0001 & 0.3466 \\
        EngagementSegment\_Inactive & 0 & 0.0009 & $-0.211$ \\
        A14Region\_APAC & 0 & 0.0014 & $-1$ \\
        AppCategoryCohort\_Productivity & 0 & 0 & $-1$ \\
        A14Region\_France & 0 & 0.0009 & $-1$ \\
        AppCategoryCohort\_General & 0 & 0.0009 & $-1$ \\
        EngagementSegment\_Low Engaged & 0 & 0.0008 & $-1$ \\
        A14Region\_Japan & 0 & 0.0008 & $-1$ \\
        A14Region\_United States & 0 & 0.0008 & $-1$ \\
        A14Region\_Western Europe & 0 & 0.0011 & $-1$ \\
        \bottomrule
        \end{tabular}
    \end{subtable}
    \hfill
    \hspace*{0.035\textwidth} 
    \hfill
    \begin{subtable}[t]{0.44\textwidth}
        \centering
        \scriptsize
        \caption{Targeted outcome}
        \begin{tabular}{lccc} 
        \toprule
         & $C_Y$ & $C_D$ & $\rho$ \\
        \midrule
        Manufacturer\_5 & 0.0032 & 0 & $-1$ \\
        Total\_Device\_Usage & 0.0016 & 0.013 & $-0.8427$ \\
        Manufacturer\_1 & 0.001 & 0 & 1 \\
        Device\_Spec\_A & 0.0006 & 0.0013 & $-0.0194$ \\
        A14Region\_Canada & 0.0006 & 0.0011 & $-0.2316$ \\
        A14Region\_CEE & 0.0004 & 0.0001 & $-1$ \\
        Device\_Spec\_B\_1 & 0.0003 & 0.0003 & $-0.0417$ \\
        A14Region\_UK & 0.0002 & 0.0008 & $-0.6281$ \\
        AppCategoryCohort\_Productivity & 0.0002 & 0 & 1 \\
        Device\_Spec\_C\_2 & 0.0002 & 0.0033 & $-0.5695$ \\
        Manufacturer\_4 & 0.0002 & 0.0011 & $-0.1681$ \\
        Device\_Spec\_C\_3 & 0.0002 & 0.0012 & $-0.2996$ \\
        A14Region\_Japan & 0.0002 & 0.0008 & $-0.3434$ \\
        Device\_Spec\_C\_1 & 0.0002 & 0.0008 & $-0.5457$ \\
        Manufacturer\_3 & 0.0002 & 0.0008 & $-0.132$ \\
        A14Region\_Korea & 0.0002 & 0.0011 & $-0.1916$ \\
        Manufacturer\_6 & 0.0001 & 0 & $-1$ \\
        Other\_Browser\_Usage & 0.0001 & 0.001 & $-1$ \\
        Browser\_2\_Usage & 0.0001 & 0 & 1 \\
        A14Region\_Greater China & 0.0001 & 0.0006 & $-0.4853$ \\
        EngagementSegment\_General & 0.0001 & 0.0014 & $-0.4892$ \\
        Device\_Spec\_B\_2 & 0.0001 & 0.0005 & $-0.4331$ \\
        A14Region\_Western Europe & 0.0001 & 0.0011 & $-0.2027$ \\
        A14Region\_Latam & 0.0001 & 0.0012 & $-0.5014$ \\
        A14Region\_India & 0.0001 & 0.001 & $-0.544$ \\
        A14Region\_United States & 0.0001 & 0.0008 & $-0.5213$ \\
        AppCategoryCohort\_General & 0.0001 & 0.0009 & $-0.2546$ \\
        AppCategoryCohort\_Media & 0.0001 & 0 & $-1$ \\
        AppCategoryCohort\_Gamer & 0.0001 & 0 & $-1$ \\
        EngagementSegment\_Inactive & 0.0001 & 0.0009 & $-0.4463$ \\
        A14Region\_France & 0.0001 & 0.0009 & $-1$ \\
        EngagementSegment\_Low Engaged & 0.0001 & 0.0008 & $-0.4973$ \\
        A14Region\_APAC & 0.0001 & 0.0014 & $-0.3106$ \\
        A14Region\_MEA & 0.0001 & 0.001 & $-0.5491$ \\
        A14Region\_Germany & 0.0001 & 0.0011 & $-0.318$ \\
        EngagementSegment\_Highly Engaged & 0.0001 & 0.0151 & $-0.6871$ \\
        AppCategoryCohort\_Developer & 0.0001 & 0 & $-1$ \\
        Browser\_1\_Usage & 0.0001 & 0 & $-1$ \\
        Browser\_4\_Usage & 0 & 0.0006 & $-0.8065$ \\
        Browser\_3\_Usage & 0 & 0.0001 & $-1$ \\
        Manufacturer\_2 & 0 & 0.0009 & $-1$ \\
        \bottomrule
        \end{tabular}
    \end{subtable}

    \vspace{0.8em}

    \caption{Covariate benchmarking for sensitivity analysis for both outcomes, sorted by $C_Y$. Each panel reports $(C_Y, C_D, \rho)$ for a single omitted observed covariate, constructed by re-estimating the model without that covariate. These benchmarks show that the \emph{observed} covariates rarely achieve both large $C_Y$ and large $C_D$, and for those that do, their empirical $\rho$’s are far from $\pm 0.5$.}
    \label{tab:covariate-benchmarking}
\end{table}

\subsection{Implementation of Sensitivity Analysis}
\label{app:sensitivity-implementation}

Let $X$ be observed covariates and $U$ a latent confounder, Following \citet{chernozhukov2022}, we define the \textit{long} regression that conditions on $W:=(X,U)$ and the \textit{short} regression that conditions on $X$ only. Denote by $\alpha(\cdot)$ the Riesz representer of the doubly robust ATE functional, $\alpha(X)=\frac{D}{e(X)}-\frac{1-D}{1-e(X)}.$ For convenience, we also define the long and short outcome regressions:  
$$
\mu_L(X)=\mu_{1,L}(X)D+\mu_{0,L}(X)(1-D),\qquad
\mu_S(X)=\mu_{1,S}(X)D+\mu_{0,S}(X)(1-D)
$$
Likewise, $\alpha_L(X)$ and $\alpha_S(X)$ are built from $ e_L(X)$ and $ e_S(X)$. Our bias bound is 

$$
|\theta_S-\theta|\;\le\;|\rho|\,C_Y\,C_D\,S,
\qquad
S^2 := \mathbb{E}\!\big[(Y-\mu_S)^2\big]\;\mathbb{E}\!\big[\alpha_S^2\big],
$$

with

$$
C_Y^2 := \frac{\mathbb{E}\!\big[(\mu_L-\mu_S)^2\big]}{\mathbb{E}\!\big[(Y-\mu_S)^2\big]},\qquad
C_D^2 := \frac{\mathbb{E}[\alpha_L^2]-\mathbb{E}[\alpha_S^2]}{\mathbb{E}[\alpha_S^2]},\qquad
\rho := \mathrm{Corr}(\mu_L-\mu_S,\ \alpha_L-\alpha_S).
$$

For empirical implementation, we estimate the long and short models with our cross-fit estimates, and use them as plug-in estimators for their population quantities: 
\[
\hat\sigma_S^2=\frac{1}{n}\sum_{i}\bigl(Y_i-\hat\mu_S(W_i)\bigr)^2,
\qquad
\hat\nu_S^2=\frac{1}{n}\sum_{i}\hat\alpha_S(W_i)^2,
\quad\text{with}\quad
\hat\alpha_S(W)=\frac{D}{\hat e_S(X)}-\frac{1-D}{1-\hat e_S(X)}.
\]
Analogously obtain $\hat\sigma_L^2,\hat\nu_L^2$ from the long models. We then plug-in these quantities to estimate our benchmark  strengths:\footnote{Our implementation follows the \texttt{DoubleML} benchmarking routine, which reports
\[
{C}_{Y,\text{DoubleML}}^2 = \frac{\hat\sigma_S^2 - \hat\sigma_L^2}{\hat\sigma_L^2}.
\]
This differs slightly from the parametrization in \citet{chernozhukov2022}, where
\[
C_Y^2 = \frac{\hat\sigma_S^2 - \hat\sigma_L^2}{\hat\sigma_S^2}.
\]
The two are related by the monotone transform $C_Y^2 = {C}_{Y,\text{DoubleML}}^2/(1+{C}_{Y,\text{DoubleML}}^2)$, so the empirical interpretation of these benchmark strengths and robustness values remains unchanged.}
\[
\widehat{C_Y^2}=\frac{\hat\sigma_S^2-\hat\sigma_L^2}{\hat\sigma_S^2},\qquad
\widehat{C_D^2}=\frac{\hat\nu_L^2-\hat\nu_S^2}{\hat\nu_S^2}.
\]

We also compute the observed benchmark set bias and its implied adversity:
\[
\widehat{\Delta\theta}=\hat\theta_S-\hat\theta_L,\qquad
\hat\rho=\mathrm{sign}(\widehat{\Delta\theta})
\!\left(
\frac{|\widehat{\Delta\theta}|}
{\sqrt{(\hat\sigma_S^2-\hat\sigma_L^2)\,(\hat\nu_L^2-\hat\nu_S^2)}}
\right).
\]

Given a $\rho$, the minimal joint strength $C_YC_D$ required to move the point estimate to zero is summarized by the robustness value:
\[
\mathrm{RV}(\rho)=\frac{|\hat\theta|}{|\rho|\,\hat S},
\qquad
\hat S=\sqrt{\hat\sigma_S^2\,\hat\nu_S^2}.
\]
At confidence level $1-\alpha$, the CI-robustness value is
\[
\mathrm{RV}_\alpha(\rho)=
\frac{\max\{\,|\hat\theta|-z_{1-\alpha}\,\widehat{\operatorname{se}}(\hat\theta),\,0\}}
{|\rho|\,\hat S}.
\]

\end{document}